\documentclass[a4paper,10pt,reqno]{amsart}

\usepackage[numbers]{natbib}
\usepackage{amsmath,amssymb,amsthm}
\usepackage[foot]{amsaddr}
\usepackage{mathtools}
\usepackage[latin1]{inputenc}
\usepackage{mathrsfs}
\usepackage{enumerate}
\usepackage{textcmds}
\usepackage[pdfusetitle]{hyperref}

\newtheorem{Theorem}{Theorem}[section]
\newtheorem{Lemma}[Theorem]{Lemma}
\theoremstyle{remark}

\newenvironment{Proof}{\begin{proof}}{\end{proof}}

\newcommand{\D}[2]{\frac{\mathrm{d}#1}{\mathrm{d}#2}}
\newcommand{\dx}{\,\mathrm{d}x}
\newcommand{\dt}{\,\mathrm{d}t}
\newcommand{\A}{^\ast}
\newcommand{\T}{^{\operatorname{T}}}
\newcommand{\pD}[2]{\frac{\partial#1}{\partial#2}}

\newcommand{\DE}{\mathcal{L}}
\newcommand{\JH}{\tau}
\newcommand{\Polynom}{\operatorname{polynomial}}

\newcommand{\Dom}{U}
\newcommand{\N}{\mathbb{N}}
\newcommand{\Z}{\mathbb{Z}}
\newcommand{\R}{\mathbb{R}}
\newcommand{\C}{\mathbb{C}}
\newcommand{\Rz}{\R^{2}}

\newcommand{\Cz}{\C^{2}}

\newcommand{\Czn}{\C^{2n}}
\newcommand{\GLzR}{\mathrm{GL}_{2}(\R)}
\newcommand{\MzR}{\mathrm{M}_{2}(\R)}

\newcommand{\MzC}{\mathrm{M}_{2}(\C)}

\newcommand{\MznC}{\mathrm{M}_{2n}(\C)}

\newcommand{\Lz}{\mathrm{L^{2}}}
\newcommand{\LznW}{\mathrm{L}_W^2((a,b),\Czn)}

\newcommand{\LzW}{\mathrm{L}_W^2((a,b),\Cz)}
\newcommand{\LWz}{\mathrm{L}_W^2}

\newcommand{\ACloc}{\mathrm{AC_{loc}}((a,b),\Czn)}
\newcommand{\Tmin}{T_\mathrm{min}}
\newcommand{\Tcls}{\overline{\Tmin}}
\newcommand{\Tmax}{T_\mathrm{max}}

\newcommand{\E}{\mathrm{e}}

\newcommand{\tr}{\operatorname{tr}}

\newcommand{\imag}{\mathrm{i}}

\newcommand{\sign}{\operatorname{sign}}

\newcommand{\term}[1]{\qq{#1}}
\newcommand{\cref}[1]{\textup{(#1)}}
\newlength{\ml}
\newcommand{\ms}{\hspace*{\ml}}

\makeatletter
\renewcommand\subsection{\@startsection{subsection}{2}%
  \z@{.5\linespacing\@plus.7\linespacing}{.5\linespacing}%
  {\normalfont\scshape}}
\renewcommand\subsubsection{\@startsection{subsubsection}{3}%
  \z@{.5\linespacing\@plus.7\linespacing}{-.5em}%
  {\normalfont\scshape}}
\makeatother

\newcommand{\Head}[4]{\title{#1}\author{#2}\email{#3}\address{#4}}
\def\sep{, }
\newcommand{\Keys}[2]{\keywords{#1}\subjclass{#2}}
\newcommand{\subsep}{}

\Head{On the deformation of linear Hamiltonian systems}{Harald Schmid}{h.schmid@oth-aw.de}{University of Applied Sciences Amberg-Weiden, Amberg, Germany}
\Keys{linear Hamiltonian systems\sep complementary triangular form\sep matrix Lax pair\sep Chandrasekhar-Page equation\sep confluent Heun equation}{34L15, 35F20, 15A21}
\thanks{\copyright\ 2021. This manuscript version is made available under the CC-BY-NC-ND 4.0 license\\ http://creativecommons.org/licenses/by-nc-nd/4.0/}

\begin{document}

\bibliographystyle{elsarticle-num}

\begin{abstract}
For linear Hamiltonian $2n\times 2n$ systems $J y'(x) = (\lambda W(x)+H(x))y(x)$ we investigate the problem how the eigenvalues $\lambda$ depend on the entries of the coefficient matrix $H$. This question turns into a deformation equation for $H$ and a partial differential equation for the eigenvalues $\lambda$. We apply our results to various examples, including generalizations of the confluent Heun equation and the Chandrasekhar-Page angular equation. We are mainly concerned with the $2\times 2$ case, and in order to reduce the degrees of freedom in $H$ as much as possible, we will first convert such systems into a complementary triangular form, which is a canonical form with a minimum number of free parameters. Furthermore, we discuss relations to monodromy preserving deformations and to matrix Lax pairs.
\end{abstract}

\maketitle

\section{Introduction}

Numerous problems in physics and technology are related to boundary eigenvalue problems. In the simpler cases these are eigenvalue problems for ordinary differential equations with Dirichlet boundary conditions, where the associated self-adjoint operators have purely discrete spectrum. Even in this situation the explicit computation of eigenvalues is quite difficult, since the solutions of an ODE, provided that they can be calculated at all, are rarely known in analytic form. However, there are certain boundary value problems where we get information about the eigenvalues without having to determine the eigenfunctions. One such example is the Chandrasekhar-Page angular equation (CPAE), which describes the angular part of the wave function of a massive particle with spin $\frac{1}{2}$ in a Kerr-Newman metric. It can be written in the form
\begin{equation} \label{CPAE}
\begin{pmatrix*}[r] 0 & 1 \\[1ex] -1 & 0 \end{pmatrix*} S'(\theta)
+ \left(\begin{array}{cc}
- \mu\cos\theta & -\frac{\kappa}{\sin\theta}-\nu\sin\theta \\[1ex]
- \frac{\kappa}{\sin\theta}-\nu\sin\theta & \mu\cos\theta \end{array}\right)S(\theta)
= \Lambda S(\theta),\quad \theta\in(0,\pi)
\end{equation}
where $\kappa\in\R\setminus(-\frac{1}{2},\frac{1}{2})$ is a fixed number, $\Lambda$ is an eigenvalue parameter, and $\mu,\nu\in\C$ are two parameters specifying the physical properties of the metric and the particle. In the Hilbert space $\Lz((0,\pi),\C^{2})$ with the scalar product $\langle S_1,S_2\rangle := \int_{0}^{\pi} S_2(\theta)^{\ast}S_1(\theta)\,\mathrm{d}\theta$ we can associate a self-adjoined operator to the left side of \eqref{CPAE}. By applying the transformation
\begin{equation*}
S(\theta) = \left(\begin{array}{cc} \sqrt{\tan\frac{\theta}{2}} & 0 \\[1ex]
0 & \sqrt{\cot\frac{\theta}{2}} \end{array}\right)y(\sin^2\tfrac{\theta}{2}),\quad
x=\sin^2\tfrac{\theta}{2}
\end{equation*}
\eqref{CPAE} is equivalent to the linear Hamiltonian system
\begin{equation} \label{CPJH}
\begin{pmatrix*}[r] 0 & 1 \\[1ex] -1 & 0 \end{pmatrix*} y'(x)
= \left(\Lambda\begin{pmatrix} \frac{1}{1-x} & 0 \\[1ex] 0 & \frac{1}{x} \end{pmatrix}
+ H(x)\right)y(x),\quad x\in(0,1)
\end{equation}
where $H(x)$ denotes the symmetrical coefficient matrix
\begin{equation*}
H(x) = H(x;\mu,\nu) = \begin{pmatrix} 
2\mu-\frac{\mu}{1-x} & 2\nu+\frac{2\kappa+1}{4x(1-x)} \\[1ex] 
2\nu+\frac{2\kappa+1}{4x(1-x)} & 2\mu - \frac{\mu}{x} \end{pmatrix}
\end{equation*}
and the square-integrability condition $\langle S,S\rangle<\infty$ becomes
\begin{equation*}
\int_{0}^{1} y(x)^\ast\left(\begin{array}{cc} \frac{1}{1-x} & 0 \\[1ex] 0 & \frac{1}{x} \end{array}\right)y(x)\dx < \infty
\end{equation*}
In \cite{BSW:2004} it is shown that the differential operator associated to \eqref{CPAE} has a purely discrete spectrum, and its simple eigenvalues $\Lambda=\Lambda_j(\mu,\nu)$ for $j\in\Z\setminus\{0\}$ depend analytically on the parameters $(\mu,\nu)$. The main result \cite[Theorem 1]{BSW:2004}, however, is the conclusion that the eigenvalues itself satisfy a quasilinear partial differential equation, namely
\begin{equation} \label{BSWE}
(\mu-2\nu\Lambda)\frac{\partial\Lambda}{\partial\mu} +
(\nu-2\mu\Lambda)\frac{\partial\Lambda}{\partial\nu} + 2\kappa\mu + 2\mu\nu = 0
\end{equation}
Using this PDE and the initial values $\Lambda_j(0,0) = \sign(j)(|\kappa|-\frac{1}{2}+|j|)$, it is comparatively simple to deduce a power series expansion for the eigenvalues $\Lambda_j(\mu,\nu)$ (see \cite[Section III]{BSW:2004}). 

The investigations in \cite{BSW:2004} revealed yet another relationship between \eqref{BSWE} and the parameter dependent system \eqref{CPJH}: Along a characteristic curve of the PDE \eqref{BSWE} certain monodromy data of \eqref{CPJH} stay constant. Such \emph{monodromy preserving deformations} of systems $\D{y}{x} = \Phi(x;t_1,\ldots,t_n)y$, whose coefficient matrix depend rationally on $x$ and on some parameters $t_k$, were extensively studied by Jimbo, Miwa, and Ueno in \cite{JMU:1981}. They were concerned with the question how the entries of $\Phi$ can be modified so that the monodromy data (Stokes multipliers, connection matrices and exponents of formal monodromy) of this meromorphic differential equation are preserved. For a system $\D{y}{x} = \Phi(x,t)y$ depending only on one parameter $t$ this problem leads to the \term{deformation equation}
\begin{equation} \label{Isomono}
\frac{\partial\Phi}{\partial t}(x,t) + \Phi(x,t)\Omega(x,t) = \Omega(x,t)\Phi(x,t) + \frac{\partial\Omega}{\partial x}(x,t)
\end{equation}
with some matrix function $\Omega(x,t)$ to be determined. Using the matrix commutator, this equation can be written in abbreviated form $\Omega_x + [\Omega,\Phi] - \Phi_t = 0$. In \cite[Section VI]{BSW:2004} it is shown that the deformation equation \eqref{Isomono} for the Chandrasekhar-Page angular equation \eqref{CPJH} is related to the characteristic curves of the PDE \eqref{BSWE}. However, there is a major difference in the methods being used: While the results in \cite{JMU:1981} were obtained by means of formal power series expansions for meromorphic differential equations, in \cite [Section III]{BSW:2004} the PDE for the eigenvalues was derived mainly with methods from analytic perturbation theory. 

The deformation equation \eqref{Isomono} also appears in the context of \emph{matrix Lax pairs}. By specifying the time evolution of the spectrum of an ordinary linear differential operator, one typically gets a nonlinear partial differential equation for the coefficients of the differential operator that depends on the location $x$ and the time $t$. In \cite{AKNS:1974} Ablowitz, Kaup, Newell and Segur developed a method which yields an integrable nonlinear PDE corresponding to a given linear differential system. Let us briefly sketch the idea of this AKNS approach: For a linear system $\D{y}{x} = \Phi(x,\lambda,t)\,y$ with some spectral parameter $\lambda$ independent from $t$ we try to find a matrix function $\Omega(x,t,\lambda)$ such that for any solution $y=y(x,t,\lambda)$ also $\pD{y}{t}-\Omega y$ satisfies this system. Thus $(y_t-\Omega y)_x = \Phi(y_t-\Omega y)$ and hence $\Phi\Omega y - (\Omega y)_x = \Phi y_t - y_{tx}$ shall be fulfilled. Provided that $y$ has continuous second partial derivatives, $y_x=\Phi y$ and $y_{tx} = y_{xt}$ imply
\begin{equation*}
(\Phi_t + \Phi\Omega - \Omega\Phi - \Omega_x)y 
= \Phi_t y + \Phi\Omega y - (\Omega y)_x = \Phi_t y + \Phi y_t - y_{tx} 
= (\Phi y)_t - y_{tx} = y_{xt}-y_{tx} = 0
\end{equation*}
and therefore $\Phi_t + \Phi\Omega - \Omega\Phi - \Omega_x = 0$. The matrix functions $\Phi$, $\Omega$ are known as AKNS pair or matrix Lax pair, and the deformation equation \eqref{Isomono} is called \term{compatibility condition} in this context. Some of these $2\times 2$ systems were studied thoroughly in \cite[section II]{AKNS:1974}, where the parameter independence of the eigenvalues (i.\,e., $\pD{\lambda}{t}\equiv 0$) and the compatibility condition \eqref{Isomono} result in, for example, the Korteweg-de Vries equation (KdV), the nonlinear Schrödinger equation and the sine-Gordon equation. It should be noted that the prerequisites for spectral preserving deformations are usually weaker than for monodromy preserving deformations: The entries of $\Phi$ have to be sufficiently smooth, but not necessarily be meromorphic. Finally, it should be mentioned that \eqref{Isomono} can also be interpreted as zero-curvature-condition of a linear connection with local connection coefficients $\Phi$ and $\Omega$ (see e.\,g. \cite [Chapter I, §2]{FT:1987}).

All these relations between \eqref{BSWE}, \eqref{Isomono} and the isomonodromic or isospectral deformations suggest that the parameter dependence of the eigenvalues can be described by a PDE not only for the CPAE but also for other differential operators as well, so that the results from \cite{BSW:2004} were not a fluke. In the present paper we will generalize \cite[Theorem 1]{BSW:2004} to linear Hamiltonian $2n\times 2n$ systems of the form
\begin{equation}
J y' = \left(\lambda W(x)+H(x;u_1,\ldots,u_m)\right)y
\label{Hamilton}
\end{equation}
where $H$ depends on $x$ and on several parameters $u_1,\ldots,u_m$. The main conclusion in section \ref{sec:HamiltonSys} is that if, in addition to some technical assumptions, the coefficients satisfy the deformation equation
\begin{equation} \label{Deform}
\pD{G}{x} + (\lambda W + H)JG - GJ(\lambda W + H) = \sum_{k=1}^{m} f_k\pD{H}{u_k} + gW
\end{equation}
then the eigenvalues $\lambda=\lambda(u_1,\ldots,u_m)$ of \eqref{Hamilton} are solutions of the quasilinear partial differential equation
\begin{equation*}
\sum_{k=1}^{m} f_k(\lambda;u_1,\ldots,u_m)\pD{\lambda}{u_k} = g(\lambda;u_1,\ldots,u_m)
\end{equation*}
To prove this statement we will primarily apply methods from analytical perturbation theory. The above result is to some extent an inversion of the AKNS method, where the evolution of the spectrum yields the compatibility condition \eqref{Isomono}. By using the PDE for the eigenvalues we will see in section \ref{sec:MatLaxPair} that the deformation equation \eqref{Deform} also corresponds to the compatibility condition of an associated matrix Lax pair. With these general considerations in mind, we study in section \ref{sec:CompTriang} the special case of singular $2\times 2$ differential systems
\begin{equation*}
z'(x) = \left(p(x) A + q(x) B + C(x;u_1,\ldots,u_m)\right)z(x),\quad x\in(a,b)
\end{equation*}
which are in the limit point case at $a$ and $b$. Here the functions $p,q:(a,b)\longrightarrow(0,\infty)$ and the matrices $A,B\in\MzR$ are fixed. First, we transform such systems into a simple standard form, which is a linear Hamiltonian system in \term{complementary triangular form}
\begin{equation*}
J y'(x) = \left(\lambda W(x) +
p(x)\begin{pmatrix} 0 & \alpha \\[1ex] \alpha & \gamma \end{pmatrix} + 
q(x)\begin{pmatrix} -\gamma & \beta \\[1ex] \ms\beta & 0 \end{pmatrix} + R(x;u_1,\ldots,u_m)\right) y(x)
\end{equation*}
If we fix $\alpha$, $\beta$ and $\gamma$, then the eigenvalues $\lambda$ of the associated self-adjoint operators depend only on the parameters $(u_1,\ldots,u_m)$. Subsequently, we apply our method to various example systems. Once we have converted them into the complementary triangular form, we use the results from section \ref{sec:HamiltonSys} to obtain partial differential equations for their eigenvalues $\lambda=\lambda(u_1,\ldots,u_m)$. In section \ref{sec:RegSingular} we will focus on $2\times 2$ systems of the type
\begin{equation*}
z'(x) = \left(\tfrac{1}{x}A + \tfrac{1}{1-x}B + C(x;u_1,\ldots,u_m)\right)z(x),\quad x\in (0,1)
\end{equation*}
and their complementary triangular form. These differential systems with regular-singular points at $x=0$ and $x=1$ are generalizations of the CPAE, and they also involve the confluent Heun equation (CHE) as another special case. Finally, in section \ref{sec:NonRegular} we will study a few more examples, such as systems with irregular singularities and even a $4\times 4$ linear Hamiltonian system.

\section{Linear Hamiltonian systems depending on several parameters}
\label{sec:HamiltonSys}

In this section we investigate linear $2n\times 2n$ Hamiltonian systems of the general form
\begin{equation}
J y'(x) = \left(\lambda W(x)+H(x;u_1,\ldots,u_m)\right)y(x) \label{LHS}
\end{equation}
on an interval $(a,b)\subset\R$, $-\infty\leq a < b\leq\infty$, with some spectral parameter $\lambda\in\C$ and a coefficient matrix $H$, which depends on $x\in(a,b)$ and one or more parameters $(u_1,\ldots,u_m)\in\Dom$, where $\Dom$ is a domain in $\R^m$. Here $y'=\pD{y}{x}=y_x$ refers to the derivative of the vector function $y:(a,b)\longrightarrow\Czn$ with respect to $x$, and $J$ is the $2n\times 2n$ matrix
\begin{equation*}
J := \begin{pmatrix} O & -E_n \\[1ex] E_n & O \end{pmatrix}
\end{equation*}
where $E_n$ denotes the $n\times n$ identity matrix. $J$ is a skew symmetric matrix satisfying $J\A = -J = J^{-1}$. In our subsequent studies we need sufficiently smooth solutions, and therefore we always assume that the following condition is fulfilled:
\begin{enumerate}[(a)]
\item The matrix functions $W:(a,b)\longrightarrow\MznC$ and $H:(a,b)\times\Dom\longrightarrow\MznC$ are at least twice continuously differentiable with respect to all variables. Moreover, $W(x)\A=W(x)>0$ and $H(x;u_1,\ldots,u_m)\A=H(x;u_1,\ldots,u_m)$ hold for all $(x;u_1,\ldots,u_m)\in(a,b)\times\Dom$.
\end{enumerate}
If condition \cref{a} is valid and $\lambda:\Dom\longrightarrow\C$, $\eta:\Dom\longrightarrow\Czn$ are at least twice continuously differentiable functions, then the $2n\times 2n$ differential system \eqref{LHS} has exactly one solution $y(x;u_1,\ldots,u_m)$ satisfying the initial condition $y(\xi;u_1,\ldots,u_m)=\eta(u_1,\ldots,u_m)$ by means of the existence and uniqueness theorem; in addition, this solution is at least twice continuously differentiable with respect to $x$ and all parameters (see e.\,g. \cite[Chap. III, \S 13, Corollary in Sec. XI]{Walter:1998}). Now, let
\begin{equation*}
\LznW := \Big\{f:(a,b)\longrightarrow\Czn\ \big|\ \int_a^b f(x)\A W(x) f(x)\dx < \infty\Big\}
\end{equation*}
be the space of square-integrable vector functions with weight function $W(x)$ and the scalar product
\begin{equation*}
\langle f,g\rangle_W := \int_a^b f(x)\A W(x)\,g(x)\dx
\end{equation*}
For fixed parameters $(u_1,\ldots,u_m)\in\Dom$, 
\begin{equation*}
\JH y := W(x)^{-1}\left(J y'(x) - H(x;u_1,\ldots,u_m)y(x)\right)
\end{equation*}
defines a formally self-adjoint differential expression in $\LznW$. The maximal operator $\Tmax y := \JH y$ generated by $\tau=\tau(u_1,\ldots,u_m)$ has the domain
\begin{equation*}
D(\Tmax) := \left\{y\in\LznW\ \big|\ y\in\ACloc\mbox{ and }\JH y\in\LznW\right\}
\end{equation*}
whereas the domain of the minimal operator $\Tmin y := \JH y$ associated to $\tau$ is given by
\begin{equation*}
D(\Tmin) := \big\{y\in D(\Tmax)\ |\ y\mbox{ has compact support in }(a,b)\big\}
\end{equation*}
$\Tmin$ is symmetric, i.\,e., Hermitian and densely defined, and $\Tmin\A = \Tmax$. The closure $\Tcls$ of the minimal operator generated by $\tau$ has the domain (cf. \cite[Theorem 3.1]{SunShi:2010})
\begin{equation}\begin{split}
D(\Tcls) = \big\{\,y\in D(\Tmax)\ |\ 
& \lim_{x\to a+}y(x)\A J w(x) = 0\mbox{ and } \\
& \lim_{x\to b-}y(x)\A J w(x) = 0\mbox{ for all }w\in D(\Tmax)\,\big\} \label{DT0}
\end{split}\end{equation}
In the following we assume that the defect indices $\dim N(\Tmax\pm\imag)$ are equal. In this case, one or more self-adjoint extensions can be associated to $\Tmin$ (see \cite[Theorem 8.6, (c)]{Weidmann:1980}). If $\tau$ is a singular differential expression which is in the limit point case at both boundary points $a$ and $b$, then the closure of the minimal operator $\Tcls$ is even self-adjoined (cf. \cite[Theorem 5.4]{SunShi:2010}), and because of $\Tmax = \Tcls\A$ (see \cite[Theorem 3.9]{Weidmann:1987}), we have $\Tmax=\Tcls$. However, if $\tau$ is regular or in the limit circle case at one boundary point, then we get self-adjoint extensions $T=T(u_1,\ldots,u_m)$ of $\tau$ in $D(\Tmax)$ only by adding appropriate boundary conditions (see \cite[Theorems 5.1 -- 5.3]{SunShi:2010}), whereby their domains may also depend on $u_1,\ldots,u_m$. In the present paper we are particularly interested in the parameter dependence of the eigenvalues of $T=T(u_1,\ldots,u_m)$. Since we do not want to stay with technical details on the boundary conditions, we will proceed from the following assumptions in addition to \cref{a}:

\begin{enumerate}[(a)] \setcounter{enumi}{1}
\item For all $(u_1,\ldots,u_m)\in\Dom$, let $T=T(u_1,\ldots,u_m)$ be a self-adjoint extension of the differential expression
\begin{equation*}
\JH y := W(x)^{-1}\left(J y'(x) - H(x;u_1,\ldots,u_m)y(x)\right),\quad\mbox{where}\quad D(T)\subset D(\Tmax)
\end{equation*}
\item For all $(u_1,\ldots,u_m)\in\Dom$, let $\lambda(u_1,\ldots,u_m)$ be an eigenvalue of $T(u_1,\ldots,u_m)$ and $y(x;u_1,\ldots,u_m)$ be an associated normalized eigenfunction, i.\,e., $\langle y,y\rangle_W=1$. In addition, suppose that $\lambda:\Dom\longrightarrow\R$ is a differentiable function and that $y:(a,b)\times\Dom\longrightarrow\Czn$ is at least twice continuously differentiable.   
\end{enumerate}

It is by no means obvious that the eigenvalues and eigenfunctions depend differentiably on the parameters $u_1\ldots,u_m$. In order to verify this, one may use uniform asymptotic integration techniques (see e.\,g. \cite[Theorem 3.5]{Behncke:2001}) or methods from analytical perturbation theory (cf. \cite[Chap. VII]{Kato:1995}, for instance). In our subsequent investigations we make use of both options, and in the next section we will even provide a sufficient criterion for \cref{b}, \cref{c}. 

Now the question arises in which manner the eigenvalues $\lambda$ depend on $(u_1,\ldots,u_m)$ for a given family of Hamiltonian operators $T(u_1,\ldots,u_m)$. It is in general rather complicated to derive a formula expression for the eigenvalues. One might try to find all solutions of the differential equation $\tau y = \lambda y$ and then determine the values $\lambda$, for which there exist nontrivial solutions satisfying the boundary conditions. This method is based on the premise that the solutions of the differential equation can be formally calculated, but this is usually not the case. Therefore we take a different approach here, and we ask instead, how the coefficient matrix $H$ has to be modified so that a comparatively simple relationship between the eigenvalues and the parameters results, for example in form of a partial differential equation. An answer will be given by Theorem \ref{DefGen} listed below, provided that in addition to \cref{a} -- \cref{c} the following condition is also valid:

\begin{enumerate}[(a)] \setcounter{enumi}{3}
\item Let $G:(a,b)\times\C\times\Dom\longrightarrow\MznC$ be a differentiable matrix function and $f_k,g:\C\times\Dom\longrightarrow\C$ be scalar functions with the property that $G=G(x;\lambda;u_1,\ldots,u_m)$ and $f_k=f_k(\lambda;u_1,\ldots,u_m)$, $g=g(\lambda;u_1,\ldots,u_m)$ satisfy the following \emph{deformation equation}:
\begin{equation} \label{DefEqu}
\pD{G}{x} + (\lambda W + H)JG - GJ(\lambda W + H) = \sum_{k=1}^{m} f_k\pD{H}{u_k} + g W
\end{equation}
\end{enumerate}

We use the normalized eigenfunction $y(x;u_1\ldots,u_m)$ from \cref{c} and the functions $f_k$, $G$ from \cref{d} to implement an auxiliary function $z = z(x;u_1,\ldots,u_m)$ by
\begin{equation} \label{pDy}
z := \sum_{k=1}^{m} f_k(\lambda(u_1,\ldots,u_m);u_1,\ldots,u_m)\pD{y}{u_k} + JG(x;\lambda(u_1,\ldots,u_m);u_1,\ldots,u_m)y
\end{equation}
From $J y_x = (\lambda W+H)y$ and Schwarz's theorem it follows that
\begin{equation*}
\pD{}{x}\left(J\pD{y}{u_k}\right) = \pD{(J y_x)}{u_k} = \left(\pD{\lambda}{u_k} W + \pD{H}{u_k}\right)y + (\lambda W+H)\pD{y}{u_k}
\end{equation*}
for $k\in\{1,\ldots,m\}$. Moreover, $y_x = -J(\lambda W+H)y$ and \eqref{DefEqu} yield
\begin{equation*}
\pD{(Gy)}{x} = G_x y - GJ(\lambda W + H)y
= \left(\sum_{k=1}^{m} f_k\pD{H}{u_k} + g W\right)y - (\lambda W+H)JGy
\end{equation*}
Since $f_k$ does not depend on $x$ and $JJ=-E_{2n}$ holds, we obtain
\begin{equation*}
J z_x
= \sum_{k=1}^{m} f_k\pD{}{x}\left(J\pD{y}{u_k}\right) - \pD{(Gy)}{x}
= (\lambda W+H)z + \left(\sum_{k=1}^{m} f_k\pD{\lambda}{u_k} - g\right)Wy
\end{equation*}
Here $\lambda=\lambda(u_1,\ldots,u_m)$, and therefore also the expression in brackets on the right side
\begin{equation*}
\mu(u_1,\ldots,u_m) 
:= \sum_{k=1}^{m} f_k(\lambda;u_1,\ldots,u_m)\pD{\lambda}{u_k} - g(\lambda;u_1,\ldots,u_m)
\end{equation*}
is a function that depends only on the parameters but not on $x$. We finally get
\begin{equation} \label{JDz}
J z_x = (\lambda W+H)z + \mu Wy
\end{equation}
With this result it is now relatively easy to prove the following assertion:

\begin{Theorem} \label{DefGen} 
Suppose that the conditions \cref{a} -- \cref{d} hold. For the functions \eqref{pDy} the limits
\begin{equation*}
\omega(u_1,\ldots,u_m) :=  
\lim_{x\to b-}(y\A J z)(x;u_1,\ldots,u_m) - \lim_{x\to a+}(y\A J z)(x;u_1,\ldots,u_m)
\end{equation*}
exist, and the eigenvalues $\lambda=\lambda(u_1,\ldots,u_m)$ satisfy on $\Dom$ the first order quasilinear PDE
\begin{equation}
\sum_{k=1}^{m} f_k(\lambda;u_1,\ldots,u_m)\pD{\lambda}{u_k} = g(\lambda;u_1,\ldots,u_m)+\omega(u_1,\ldots,u_m) \label{GenPDE}
\end{equation}
\end{Theorem}

\begin{Proof} 
Multiplying \eqref{JDz} from the left by $y\A$ produces
\begin{align*}
\pD{}{x}(y\A J z) = y_x\A Jz + y\A(\lambda W+H)z + \mu y\A W y
\end{align*}
From $J\A=-J$, $W=W\A$, $H = H\A$ and \eqref{LHS} it follows that $y_x\A = y\A(\lambda W+H)J$, and therefore 
\begin{equation*} 
\pD{}{x}(y\A J z) =  \mu y\A W y
\end{equation*}
remains. Integration over some subinterval $[\alpha,\beta]\subset(a,b)$ gives
\begin{equation*}
\left(\sum_{k=1}^{m} f_k\pD{\lambda}{u_k} - g\right)\int_\alpha^\beta y(x)\A W(x)y(x)\dx
= (y\A J z)(\beta)-(y\A J z)(\alpha)
\end{equation*}
For $\alpha\to a$ and $\beta\to b$ the integral on the left converges to 
$\langle y,y\rangle_W$, and thus also the limits on the right side do exist:
\begin{equation*}
\left(\sum_{k=1}^{m} f_k\pD{\lambda}{u_k} - g\right)\langle y,y\rangle_W = \omega(u_1,\ldots,u_m)
\end{equation*}
Since $y$ is assumed to be a normalized eigenfunction according to \cref{c}, we can omit $\langle y,y\rangle_W=1$, and we finally get the partial differential equation \eqref{GenPDE}.
\end{Proof}

In order to apply Theorem \ref{DefGen}, we first have to find solutions $G$, $f_k$, $g$ of the deformation equation \eqref{DefEqu}, and then we need to determine the values $\omega(u_1,\ldots,u_m)$. If the singular differential operator is in the limit point case at both boundary points and if, in addition, $z(x;u_1,\ldots,u_m)$ are square-integrable functions on $(a,b)$, then we may already assume $\omega\equiv 0$ according to

\begin{Theorem} \label{DefL2W}
Suppose that the conditions \cref{a} -- \cref{d} are satisfied. If $\tau$ is in the limit point case at $a$ and $b$, and if the function \eqref{pDy} satisfies $z(x;u_1,\ldots,u_m)\in\LznW$ for all $(u_1,\ldots,u_n)\in\Dom$, then the eigenvalues $\lambda=\lambda(u_1,\ldots,u_m)$ are solutions of the first order quasilinear PDE
\begin{equation}
\sum_{k=1}^{m} f_k(\lambda;u_1,\ldots,u_m)\pD{\lambda}{u_k} = g(\lambda;u_1,\ldots,u_m) \label{PDE0}
\end{equation}
Moreover, $z(x;u_1,\ldots,u_m)$ itself is a solution of the differential system \eqref{LHS} for all $(u_1,\ldots,u_m)\in\Dom$.
\end{Theorem}

\begin{Proof}
For some fixed but arbitrary parameter set $(u_1,\ldots,u_n)\in\Dom$ we have
\begin{equation*}
\tau z = W^{-1}(Jz_x - Hz) = \lambda z + \mu y
\end{equation*}
due to \eqref{JDz}. If $z\in\LznW$ holds, then $y\in\LznW$ implies $\tau z\in\LznW$ and therefore $z\in D(\Tmax)$. Now, if the singular differential expression $\tau(u_1,\ldots,u_m)$ is in the limit point case at $a$ and $b$, then $T=\Tmax=\Tcls$ is the only self-adjoint extension to the minimal operator. If we apply \eqref{DT0} to $y\in D(T)=D(\Tcls)$ and $w=z\in D(\Tmax)$, then we obtain
\begin{equation*}
\lim_{x\to a+}y(x)\A J z(x) = 0\quad\mbox{and}\quad\lim_{x\to b-}y(x)\A J z(x) = 0
\end{equation*}
Consequently,
\begin{equation*}
\omega(u_1,\ldots,u_m) = \lim_{x\to b-}(y\A J z)(x) - \lim_{x\to a+}(y\A J z)(x) = 0
\end{equation*}
Inserting $\omega=0$ into \eqref{GenPDE} gives \eqref{PDE0} and $\mu = 0$. In particular, \eqref{JDz} is reduced to $J z_x = (\lambda W+H)z$. Thus $z\in D(\Tmax)=D(T)$ is also a solution of $\tau z = \lambda z$.
\end{Proof}

Under the conditions of Theorem \ref{DefL2W} not only the partial differential equation for the eigenvalues can be simplified: If $z\not\equiv 0$, then $z = \sum_{k=1}^{m} f_k\pD{y}{u_k} + JGy$ is even an eigenfunction associated with $\lambda$.

In the remainder of this section we consider Hamiltonian systems with linear dependence on the parameters. This means that the coefficient matrix depends on the parameters $u_1,\ldots,u_m$ in the following way:
\begin{equation}
H(x;u_1,\ldots,u_m) = H_0(x) + \sum_{k=1}^m u_k\cdot H_k(x) \label{LinearH}
\end{equation}
We will give a slightly different version of Theorem \ref{DefL2W}, which is merely derived with methods from analytical perturbation theory, and we will also replace \cref{a} -- \cref{d} by the subsequent assumptions that are easier to verify in practice:

\begin{enumerate}[(A)]
\item $W,H_k:(a,b)\longrightarrow\MznC$ are matrix functions of class $\mathrm{C}^2$ satisfying $W(x)\A=W(x)>0$ and $H_k(x)\A=H_k(x)$ for all $x\in(a,b)$ and $k=0,\ldots,m$. Moreover, $|W(x)^{-1/2}H_k(x)W(x)^{-1/2}|\leq C$ holds for all $x\in(a,b)$ and $k=1,\ldots,m$ with some constant $C\geq 0$.
\item $\JH_0 y := W(x)^{-1}(Jy'(x) - H_0(x)y(x))$ is a singular differential expression on $(a,b)$ which is in the limit point case at both boundary points; thus, in particular, the closure of the associated minimal operator $T_0=\Tcls$ is the only self-adjoint extension of $\JH_0$ in $\LznW$. 
\item $\lambda_0\in\R$ is a simple eigenvalue of $T_0$, i.\,e., $\lambda\in\sigma(T_0)$ is isolated with algebraic multiplicity $1$.
\item $G:(a,b)\times\C\times\C^m\longrightarrow\MznC$ is a differentiable matrix function and $f_k,g:\C\times\C^m\longrightarrow\C$ are scalar functions satisfying for all $(x;\lambda;u_1,\ldots,u_m)\in (a,b)\times\C\times\C^m$ the deformation equation
\begin{equation}
\pD{G}{x} + (\lambda W + H)JG - GJ(\lambda W + H) = \sum_{k=1}^{m} f_k H_k + g W \label{DefLin}
\end{equation}
\end{enumerate}

Before we state our result, we will briefly discuss how these conditions are related to the assumptions \cref{a} -- \cref{d} from above. Obviously, the weight function $W(x)$ and the coefficient matrix $H(x;u_1,\ldots,u_m)$ satisfy \cref{a}. Since $W(x)$ is a positive definite Hermitian matrix, for each $x\in(a,b)$ there exists a unique invertible Hermitian matrix $W(x)^{1/2}$ such that $W(x) = W(x)^{1/2}\cdot W(x)^{1/2}$. Furthermore, the multiplication operator given by $A_k y := -W(x)^{-1}H_k(x)\,y(x)$ is symmetric and bounded, since for any $y\in\LznW$ we have 
\begin{align*}
\|A_k y\|_W^2 
& = \int_a^b \left(W(x)^{-1}H_k(x)\,y(x)\right)\A W(x)\left(W(x)^{-1}H_k(x)\,y(x)\right)\dx \\
& = \int_a^b |W(x)^{-1/2}H_k(x)W(x)^{-1/2}W(x)^{1/2}y(x)|^2\dx 
\leq\int_a^b C^2|W(x)^{1/2}y(x)|^2\dx = C^2\|y\|_W^2
\end{align*}
From \cite[Chap. V, \S\,4, Theorem 4.3 and Chap. VII, \S\,2, Theorem 2.6]{Kato:1995} it follows that
\begin{equation*}
T(u_1,\ldots,u_m) := T_0 + \sum_{k=1}^m u_k\cdot A_k
\end{equation*}
defines a holomorphic family of self-adjoint operators (in the sense of Kato), where its domain $D(T)=D(T_0)$ is independent of $(u_1,\ldots,u_m)\in\C^m$. For a fixed set of real parameters $(u_1,\ldots,u_m)\in\R^m$ the operator $T(u_1,\ldots,u_m)$ is a self-adjoint extension of the minimal operator generated by $\tau(u_1,\ldots,u_m)$. Furthermore, $\lambda_0\in\R$ is a simple eigenvalue of $T_0=T(0,\ldots,0)$. As shown in \cite[Chap. VII, \S\,2, Sec. 4 and \S\,3, Sec. 1 -- 2]{Kato:1995} there exists a domain $\Dom\subset\R^m$ with $(0,\ldots,0)\in\Dom$ as well as an analytical function $\lambda=\lambda(u_1,\ldots,u_m)$ with $\lambda(0,\ldots,0)=\lambda_0$, such that $\lambda(u_1,\ldots,u_m)$ is a simple eigenvalue of $T(u_1,\ldots,u_m)$ for all $(u_1,\ldots,u_m)\in\Dom$. In the following, such a function $\lambda:\Dom\longrightarrow\R$ is called an \emph{analytical continuation} of the simple eigenvalue $\lambda_0$. According to \cite[Chap. VII, \S\,3, Sec. 4]{Kato:1995} even the normalized eigenfunctions $y(x;u_1,\ldots,u_m)$ depend analytically on $(u_1,\ldots,u_m)$, so that the self-adjoint operators $T(u_1,\ldots,u_m)$ satisfy the above prerequisites \cref{b}, \cref{c}. Finally, condition \cref{D} is a special case of \cref{d} with $\pD{H}{u_k}=H_k$.

\begin{Theorem} \label{Family}
Under the assumptions \cref{A} -- \cref{D} we obtain that
\begin{equation*}
T(u_1,\ldots,u_m) := T_0 + \sum_{k=1}^m u_k\cdot A_k,\quad A_k y := -W(x)^{-1}H_k(x)\,y(x)
\end{equation*}
with $D(T)=D(T_0)$ forms a holomorphic family of self-adjoint differential operators. Furthermore, a simple eigenvalue $\lambda_0$ of $T_0$ can be continued analytically (as simple eigenvalue) to some domain $\Dom\subset\R^m$ with $(0,\ldots,0)\in\Dom$. If, in addition, the normalized eigenfunctions $y=y(x;u_1,\ldots,u_m)$ associated with $\lambda=\lambda(u_1,\ldots,u_m)$ satisfy the boundary conditions
\begin{equation*}
\lim_{x\to a+}y(x)\A G(x;\lambda;u_1,\ldots,u_m)y(x) =
\lim_{x\to b-}y(x)\A G(x;\lambda;u_1,\ldots,u_m)y(x) 
\end{equation*}
then the eigenvalues of $T(u_1,\ldots,u_m)$ solve the partial differential equation \eqref{PDE0} on $\Dom$.
\end{Theorem}

\begin{Proof}
From \cite[Chap. VII, \S\,3, Sec. 4]{Kato:1995} it follows that the derivative of the eigenvalues with respect to the parameter $u_k$ is given by
\begin{equation*}
\pD{\lambda}{u_k} 
= \left\langle\pD{T}{u_k} y,y\right\rangle_W = \left\langle A_k y,y\right\rangle_W 
= \langle -W^{-1}H_k y,y\rangle_W = -\int_a^b y(x)\A H_k(x)y(x)\dx
\end{equation*}
Since the functions $f_k=f_k(\lambda;u_1,\ldots,u_m)$ are independent of $x$, we obtain
\begin{equation*}
\sum_{k=1}^m f_k(\lambda;u_1,\ldots,u_m)\pD{\lambda}{u_k} 
= -\int_a^b y(x)\A\left(\sum_{k=1}^{m} f_k H_k(x)\right)y(x)\dx
\end{equation*}
and from the deformation equation \eqref{DefLin} it follows that
\begin{align*}
\pD{(y\A Gy)}{x}
& = y_x\A G y + y\A G_x y + y\A G y_x = y\A(\lambda W + H)JG y + y\A G_x y - y\A GJ(\lambda W + H)y \\
& = y\A\left(\sum_{k=1}^{m} f_k H_k + g W\right)y
\end{align*}
As $g=g(\lambda;u_1,\ldots,u_m)$ is also not dependent on $x$, and because $\langle y,y\rangle_W = 1$ holds, we get
\begin{align*}
\int_a^b\pD{}{x}(y\A G y)\dx
& = \int_a^b y\A\left(\sum_{k=1}^{m} f_k H_k\right)y\dx + g\int_a^b y\A Wy\dx \\
& = -\sum_{k=1}^m f_k(\lambda;u_1,\ldots,u_m)\pD{\lambda}{u_k} + g(\lambda;u_1,\ldots,u_m)
\end{align*}
Due to the boundary conditions, the eigenfunctions satisfy
\begin{equation*}
\int_a^b\pD{}{x}(y\A G y)\dx 
= \lim_{x\to b-}y(x)\A G(x;\lambda;u_1,\ldots,u_m)y(x)
- \lim_{x\to a+}y(x)\A G(x;\lambda;u_1,\ldots,u_m)y(x) = 0
\end{equation*}
and finally we again arrive at the partial differential equation \eqref{PDE0} for the eigenvalues.
\end{Proof}

Theorem \ref{Family} can be obtained in a quite different way, too, namely by means of Theorem \ref{DefGen}. As already mentioned, the assumptions \cref{A} -- \cref{D} result in the conditions \cref{a} -- \cref{d}, and for a holomorphic family of self-adjoint operators the derivatives $\pD{y}{u_k}$ of the normalized eigenfunctions are also elements of $D(T_0)$, so that according to \eqref{DT0} the limits
\begin{equation*}
\lim_{x\to b-}y(x)\A J\pD{y}{u_k}(x) = \lim_{x\to a+}y(x)\A J\pD{y}{u_k}(x) = 0
\end{equation*}
exist. In combination with $\lim_{x\to b-}y(x)\A G(x)y(x) = \lim_{x\to a+}y(x)\A G(x)y(x)$ we receive
\begin{equation*}
\lim_{x\to b-} z(x;u_1,\ldots,u_m) = \lim_{x\to a+} z(x;u_1,\ldots,u_m)
\end{equation*}
for the functions \eqref{pDy}, and this implies $\omega(u_1,\ldots,u_m)\equiv 0$ in \eqref{GenPDE}.

\section{Some notes on the deformation equation and the PDE for the eigenvalues}
\label{sec:MatLaxPair}

In order to get a PDE for the eigenvalues of \eqref{LHS}, we first have to solve the deformation equation \eqref{DefEqu}, where the scalar functions $f_k(\lambda;u_1,\ldots,u_m)$ and $g(\lambda;u_1,\ldots,u_m)$ can be specified as required. Once a solution $G(x)=G(x;\lambda;u_1,\ldots,u_m)$ of \eqref{DefEqu} is known, we have to verify $z = \sum_{k=1}^{m} f_k\pD{y}{u_k} + JGy\in\LzW$ to apply Theorem \ref{DefL2W} or $\lim_{x\to a+}y(x)\A G(x)y(x) = \lim_{x\to b-}y(x)\A G(x)y(x)$ to apply Theorem \ref{Family}. In any case, it is necessary to study the asymptotic behavior of $G(x)$ at the boundary points $a$ and $b$. For this purpose we bring the differential system \eqref{LHS} to the form
\begin{equation*}
y'(x) = \Phi(x;\lambda;u_1,\ldots,u_m)y(x),\quad x\in(a,b)
\end{equation*}
with the coefficient matrix
\begin{equation*}
\Phi(x;\lambda;u_1,\ldots,u_m) = -J\big(\lambda W(x) + H(x;u_1,\ldots,u_m)\big)
\end{equation*}
If $\DE$ denotes the linear differential operator
\begin{equation*}
\DE := \sum_{k=1}^{m} f_k(\lambda;u_1,\ldots,u_m)\pD{}{u_k} + g(\lambda;u_1,\ldots,u_m)\pD{}{\lambda}
\end{equation*}
and $\Omega := -JG$, then after multiplication with $J$ from the left the deformation equation \eqref{DefEqu} becomes
\begin{equation*}
\DE\Phi + \Phi\Omega - \Omega\Phi - \pD{\Omega}{x} = 0
\end{equation*}
or briefly $\DE\Phi + [\Phi,\Omega] - \Omega_x = 0$. Now we will prove that the solutions of \eqref{DefEqu} can be calculated using a (sufficiently smooth) fundamental matrix of \eqref{LHS}.

\begin{Theorem} \label{Fundamental}
Let $Y=Y(x;\lambda;u_1,\ldots,u_m)$ be fundamental matrix of \eqref{LHS} which is twice continuously differentiable on $(a,b)\times\C\times\Dom$. A matrix function $G:(a,b)\times\C\times\Dom\longrightarrow\MznC$ is a solution of the deformation equation \eqref{DefEqu} if and only if
\begin{equation*}
G(x;\lambda;u_1,\ldots,u_m) = J(\DE Y-YC)Y^{-1}
\end{equation*}
where the matrix $C=C(\lambda;u_1,\ldots,u_m)\in\MznC$ is independent of $x$.
\end{Theorem}

\begin{Proof} 
We define $Z:=\DE Y - \Omega Y$, where $\Omega=\Omega(x;u_1,\ldots,u_m)$ is an arbitrary $2n\times 2n$ matrix function which is differentiable with respect to $x$. The fundamental matrix $Y$ satisfies $Y_x = \Phi Y$, and Schwarz's theorem $(\DE Y)_x=\DE(Y_x)=\DE(\Phi Y)$ implies
\begin{equation*}
Z_x - \Phi Z
= \DE(\Phi Y) - \Omega_x Y - \Omega Y_x - \Phi\DE Y + \Phi\Omega Y 
= (\DE\Phi) Y - \Omega_x Y - \Omega\Phi Y + \Phi\Omega Y
\end{equation*}
or
\begin{equation}
Z_x - \Phi Z = (\DE\Phi + [\Phi,\Omega] - \Omega_x)Y \label{DefZ}
\end{equation}
Now, if $G$ satisfies \eqref{DefEqu}, then $\Omega:=-JG$ solves the equation $\DE\Phi + [\Phi,\Omega] - \Omega_x = 0$, and the right side of \eqref{DefZ} vanishes. Thus $Z_x = \Phi Z$, and therefore we get $\DE Y - \Omega Y = Z = YC$ with some matrix function $C$ independent of $x$. If conversely $Z := \DE Y - \Omega Y = YC$ with a matrix function $C:\C\times\Dom\longrightarrow\MznC$, then in \eqref{DefZ} the expression on the left side becomes $Z_x - \Phi Z = 0$. As $Y$ is invertible, it follows that $\DE\Phi + [\Phi,\Omega] - \Omega_x = 0$, and therefore $G$ is a solution of \eqref{DefEqu}. Hence, \eqref{DefEqu} is valid if and only if $\DE Y + JGY=Z=YC$ holds, and that means $G = J(\DE Y-YC)Y^{-1}$.
\end{Proof}

A sufficiently smooth fundamental matrix as required by Theorem \ref{Fundamental} can always be found, provided that one of the conditions \cref{a} or \cref{A} is fulfilled. According to the existence and uniqueness theorem the fundamental matrix with initial value $Y(\xi;\lambda;u_1,\ldots,u_m)=E_{2n}$ at some fixed location $\xi\in(a,b)$ depends twice continuously differentiable on all variables. However, the explicit calculation of such a fundamental matrix $Y(x;\lambda;u_1,\ldots,u_m)$ is quite difficult. Nevertheless, Theorem \ref{Fundamental} is not only of theoretical interest: The practical use for the deformation equation is to determine the formal structure of its solutions, which in turn allows us to examine the asymptotic behavior of $G(x)$ at the boundary points $a$ and $b$. For example, if we can convert the differential system \eqref{LHS} into Levinson's form and thus obtain a fundamental matrix of the form $Y(x) = P(x)\E^{\Lambda(x)}$ with some diagonal matrix function $\Lambda=\Lambda(x;\lambda;u_1,\ldots,u_m)$, then the solutions of \eqref{DefEqu} take the form
$G = J\big(\DE P + P\DE\Lambda - P\,\E^{\Lambda}C\E^{-\Lambda}\big)P^{-1}$.

In principle, one can solve not only the deformation equation but also the quasilinear partial differential equation \eqref{PDE0} for the eigenvalues, namely by the method of characteristics. A characteristic curve to \eqref{PDE0} with curve parameter $t\in(\alpha,\beta)\subset\R$ is described by the nonlinear autonomous differential system
\begin{align*}
\D{u_k}{t} & = f_k(\lambda;u_1,\ldots,u_m) \quad (k=1,\ldots,m) \\
\D{\lambda}{t} & = g(\lambda;u_1,\ldots,u_m)
\end{align*}
Along such a characteristic curve $u_1(t),\ldots,u_m(t),\lambda(t)$ we define the matrix functions
\begin{align*}
  \Phi(x,t) & := -J\big(\lambda(t)W(x) + H(x;u_1(t),\ldots,u_m(t))\big) \\
\Omega(x,t) & := -J G(x;\lambda(t);u_1(t),\ldots,u_m(t))
\end{align*}
Multiplying the deformation equation \eqref{DefEqu} from the left by $-J$ yields
\begin{equation*}
-J\pD{G}{x} - J(\lambda W+H)JG + JGJ(\lambda W + H) = -J\left(g W + \sum_{k=1}^m f_k\pD{H}{u_k}\right)
\end{equation*}
If we replace $\Omega=-JG$ and $\Phi=-J(\lambda W+H)$ on the left side, then we get
\begin{equation*}
\pD{\Omega}{x} - \Phi\Omega + \Omega\Phi = -J\left(g W + \sum_{k=1}^m f_k\pD{H}{u_k}\right)
\end{equation*}
On the other hand,
\begin{equation*}
\pD{\Phi}{t} 
= -J\left(\D{\lambda}{t}\,W + \sum_{k=1}^m\D{u_k}{t}\pD{H}{u_k}\right) 
= -J\left(g W + \sum_{k=1}^m f_k\pD{H}{u_k}\right)
\end{equation*}
Finally, we arrive at
\begin{equation*}
\pD{\Omega}{x} - \Phi\Omega + \Omega\Phi = \pD{\Phi}{t}
\end{equation*}
or briefly $\Omega_x + [\Omega,\Phi] - \Phi_t = 0$, and this is the compatibility condition for the linear systems
\begin{equation*}
y_x = \Phi(x,t)y,\quad y_t = \Omega(x,t)y
\end{equation*}
(cf. \cite[Sec 1.2]{AS:1981}). Generally speaking, along a characteristic curve of the PDE \eqref{PDE0} the deformation equation \eqref{DefEqu} can be interpreted as compatibility condition for the matrix Lax pair
\begin{align*}
J y_x & = \big(\lambda(t) W(x) + H(x;u_1(t),\ldots,u_m(t))\big)y \\
J y_t & = G(x;\lambda(t);u_1(t),\ldots,u_m(t))y
\end{align*}

\section{Linear $2\times 2$ differential systems in complementary triangular form}
\label{sec:CompTriang}

In this section we consider linear $2\times 2$ systems of the type
\begin{equation}
z'(x) = \left(p(x) A + q(x) B + C(x)\right)z(x),\quad x\in (a,b) \label{GDS}
\end{equation}
on an open interval $(a,b)\subset\R$, $-\infty\leq a < b\leq\infty$ with some real matrices $A,B\in\MzR$, where the functions $p,q:(a,b)\longrightarrow(0,\infty)$ and $C:(a,b)\longrightarrow\MzR$ are supposed to be locally integrable. In this general form, each of the coefficient matrices $A$ and $B$ already has four adjustable entries, and in addition, \eqref{GDS} is not yet formulated as an eigenvalue problem. Our next goal is to transform the differential system \eqref{GDS} into a Hamiltonian system of type \eqref{LHS}. Since each entry in $H$ may become a variable in the PDE \eqref{PDE0}, which will be more and more difficult to solve as the number of variables increases, we further try to minimize the number of free parameters in $H$. As a first step in this direction we may assume without restriction that $\tr A = \tr B = 0$ holds and that $\tr C(x)=0$ is valid for all $x\in(a,b)$. This can be arranged as follows: If we define the function
\begin{equation*}
w(x) := \exp\left(\tfrac{1}{2}\int_c^x p(t)\tr A + q(t)\tr B + \tr C(t)\dt\right)
\end{equation*}
with some fixed but arbitrary point $c\in(a,b)$, then the transformation $z(x) = w(x)\tilde z(x)$ gives the $2\times 2$ differential system
\begin{equation*}
\tilde z'(x) = \left(p(x)\tilde A + q(x)\tilde B + \tilde C(x)\right)\tilde z(x)
\end{equation*}
where 
\begin{equation*}
\tilde A = A - \tfrac{\tr A}{2}E_2,\quad
\tilde B = B - \tfrac{\tr B}{2}E_2,\quad
\tilde C(x) = C(x) - \tfrac{\tr C(x)}{2}E_2
\end{equation*} 
(here $E_2$ denotes the $2\times 2$ identity matrix). These matrices satisfy $\tr\tilde A = \tr\tilde B = 0$ and $\tr\tilde C(x)\equiv 0$, as required. Since we want to associate a self-adjoint differential operator to the Hamiltonian system, preferably without imposing additional boundary conditions, we will in addition assume the limit point case at both boundary points $a$ and $b$. To this end, we suppose that the $2\times 2$ matrices $A$, $B$ each have two different real eigenvalues. In the case $\tr A = \tr B = 0$ this is equivalent to $\det A < 0$ and $\det B < 0$. For a further reduction of the free entries in $A$ and $B$ we will also benefit from the condition $\det[A,B]\neq 0$, where $[A,B] := AB-BA$ denotes the matrix commutator. In summary, we start from the following prerequisites:
\begin{equation}
\tr A = \tr B = 0,\quad\det A < 0,\quad\det B < 0,\quad\det[A,B]\neq 0 \label{CTM}
\end{equation}

\begin{Lemma}\label{MatCTF} Suppose that $A,B\in\MzR$ satisfy condition \eqref{CTM}. If we define $\alpha := \sqrt{-\det A}$ and $\beta := \sqrt{-\det B}$, then there exist numbers $\gamma,\lambda\in\R$ and an invertible matrix $T\in\MzR$ with the property
\begin{equation} \label{CTF}
T^{-1}A\,T = \begin{pmatrix*} \alpha & \gamma+\lambda \\[1ex] 0 & -\alpha \end{pmatrix*},\quad
T^{-1}B\,T = \begin{pmatrix*} \beta & \ms 0 \\[1ex] \gamma-\lambda & -\beta \end{pmatrix*}
\end{equation}
Moreover, $T\in\GLzR$ can be adjusted such that $\gamma = 0$ for the case $\det(\alpha B-\beta A)<0$, $\lambda = 0$ if $\det(\alpha B-\beta A)>0$, and $\lambda=\pm\gamma$ holds in case of $\det(\alpha B-\beta A)=0$.
\end{Lemma}

\begin{Proof} Because of $\tr A = \tr B = 0$, the eigenvalues of $A$ and $B$ are $\pm\alpha$ and $\pm\beta$, respectively. Moreover, from $\det[A,B]\neq 0$ it follows that the matrices $A,B\in\MzR$ do not have common eigenvectors (cf. \cite[Theorem 3.1]{Shemesh:1984}). Hence, if $u_1\in\Rz$ is an eigenvector of $A$ associated to the eigenvalue $\alpha$ and $u_2\in\Rz$ is an eigenvector of $B$ associated to the eigenvalue $-\beta$, then $u_1$, $u_2$ are linearly independent, and for the invertible matrix $U := \begin{pmatrix} u_1\,|\,u_2 \end{pmatrix}\in\MzR$ we get
\begin{equation*}
\hat A := U^{-1}A\,U = \begin{pmatrix} \alpha & \ms c \\[1ex] 0 & -\alpha \end{pmatrix},\quad
\hat B := U^{-1}B\,U = \begin{pmatrix} \beta  & \ms 0 \\[1ex] d & -\beta  \end{pmatrix}
\end{equation*}
with certain numbers $c,d\in\R$. This result corresponds to \eqref{CTF} with $T:=U$ and $\gamma := (c+d)/2$, $\lambda := (c-d)/2$. In addition, $\det(\alpha B-\beta A) = \det(\alpha\hat B-\beta\hat A) = \alpha\beta cd$. Since $\alpha>0$ and $\beta>0$, $\det(\alpha B-\beta A)=0$ implies $c=0$ or $d=0$, and thus $\lambda=-\gamma$ or $\lambda=\gamma$. Furthermore, if $s\neq 0$, then
\begin{equation*}
S^{-1}\hat A\,S = \begin{pmatrix*} \alpha & cs \\[1ex] 0 & -\alpha \end{pmatrix*}
\quad\mbox{and}\quad
S^{-1}\hat B\,S = \begin{pmatrix*} b & \ms 0 \\[1ex] \frac{d}{s} & -b \end{pmatrix*}
\quad\mbox{with}\quad
S := \begin{pmatrix} 1 & 0 \\[1ex] 0 & s \end{pmatrix}
\end{equation*}
In the case $\det(\alpha B-\beta A)<0$ we obtain $cd<0$. If we set $\lambda:=\sqrt{-cd}$ and $s := \lambda/c = -d/\lambda$, then $T := US$ gives \eqref{CTF} with $\gamma=0$. On the other hand, $\det(\alpha B-\beta A)>0$ implies $cd>0$. Here $\gamma:=\sqrt{cd}$, $s := \gamma/c = d/\gamma$ and $T := US$ result in \eqref{CTF} with $\lambda=0$.
\end{Proof}

In Lemma \ref{MatCTF} the matrices $A$ and $B$ are converted to a so-called \emph{complementary triangular form}. Such a simultaneous reduction to complementary triangular matrices can already be realized under weaker conditions on $A$ and $B$, see \cite{Bart:1988}. With the condition \eqref{CTM}, however, we can additionally arrange the entries on the matrix diagonal, and via the sign of $\det(\alpha B-\beta A)$ we can also configure the off-diagonal elements. After a transformation to the complementary triangular form \eqref{CTF} there are only three adjustable entries left, namely the two positive eigenvalues $\alpha$, $\beta$ and the entries $\gamma$ or $\lambda$, one of which is zero or both have the same absolute value.

\begin{Theorem}\label{SysCTF} Suppose that $A,B\in\MzR$ satisfy the condition \eqref{CTM} and that $C:(a,b)\longrightarrow\MzR$ fulfills $\tr C(x)\equiv 0$ on $(a,b)$. Then there exist an invertible matrix $T\in\MzR$ and real numbers $\alpha,\beta>0$, $\gamma$ and $\lambda$, such that by means of the transformation $z(x) = T y(x)$ the linear differential system \eqref{GDS} is equivalent to
\begin{equation}
J y'(x) = \left(\lambda W(x)+
p(x)\begin{pmatrix} 0 & \alpha \\[1ex] \alpha & \gamma \end{pmatrix} + 
q(x)\begin{pmatrix} -\gamma & \beta \\[1ex] \ms\beta & 0 \end{pmatrix} + R(x)\right) y(x) \label{CTHS}
\end{equation}
where
\begin{equation*}
J := \begin{pmatrix} 0 & -1 \\[1ex] 1 & \ms 0 \end{pmatrix},\quad
W(x) := \begin{pmatrix} q(x) & 0 \\[1ex] 0 & p(x) \end{pmatrix},
\quad R(x) = R(x)\A
\end{equation*}
for all $x\in(a,b)$. Subsequently, the linear Hamiltonian system \eqref{CTHS} will be called the complementary triangular form of \eqref{GDS}.  
\end{Theorem}

\begin{Proof} We take the matrix $T$ from Lemma \ref{MatCTF}. The transformation $z(x) = Ty(x)$ leads to the system  $y'(x) = \big(p(x)\hat A + q(x)\hat B + \hat C(x)\big)y(x)$ with 
\begin{gather*}
\hat A := T^{-1}A\,T = \begin{pmatrix} \alpha & \gamma+\lambda \\[1ex] 0 & -\alpha \end{pmatrix},\quad
\hat B := T^{-1}A\,T = \begin{pmatrix} \beta & \ms 0 \\[1ex] \gamma-\lambda & -\beta \end{pmatrix} \\
\hat C(x) := T^{-1}C(x)\,T = \begin{pmatrix} c_2(x) & \ms c_3(x) \\[1ex] c_1(x) & -c_2(x) \end{pmatrix}
\end{gather*}
and certain locally integrable functions $c_k:(a,b)\longrightarrow\R$, where $\tr\hat C(x) = \tr C(x) = 0$ holds for all $x\in(a,b)$. After multiplication from the left by the matrix $J$ this system is equivalent to
\begin{equation*}
J y'(x) = \left(p(x)\,J\hat A + q(x)\,J\hat B + J\hat C(x)\right) y
\end{equation*}
which complies with \eqref{CTHS}, where $R(x) := J\hat C(x)$ is a symmetric matrix function on $(a,b)$.
\end{Proof}

In the following we assume that the $2\times 2$ differential system has already been transformed to the complementary triangular form. If necessary, we can convert such a linear Hamiltonian system \eqref{CTHS} into the \term{normal form}
\begin{equation}
y'(x) = \left(
p(x)\begin{pmatrix} \alpha & \gamma+\lambda \\[1ex] 0 & -\alpha \end{pmatrix} + 
q(x)\begin{pmatrix} \beta & \ms 0 \\[1ex] \gamma-\lambda & -\beta \end{pmatrix} - JR(x)\right) y(x) \label{CTDS}
\end{equation}
or even write it as an eigenvalue equation $\tau y = \lambda y$ with the differential expression
\begin{equation}
\tau y := W(x)^{-1}\left(J y' - \left(p(x)\begin{pmatrix} 0 & \alpha \\[1ex] \alpha & \gamma \end{pmatrix} + q(x)\begin{pmatrix} -\gamma & \beta \\[1ex] \ms\beta & 0 \end{pmatrix} + R(x)\right)y\right) \label{CTDE}
\end{equation}
Now that we have found a canonical form \eqref{CTHS} for our $2\times 2$ differential systems, we are going to investigate the asymptotic behavior of their solutions at the boundary points $a$ and $b$ with regard to self-adjoint extensions of $\tau$. For convenience, we aim at the limit point case at both boundary points, and therefore we will proceed from the following assumptions on the coefficients:
\begin{enumerate}[(i)]
\item $p,q:(a,b)\longrightarrow(0,\infty)$ are locally integrable functions satisfying
\begin{gather*}
\int_x^c p(t)\dt \to\infty\quad(x\to a),\quad \int_c^b p(t)\dt < \infty \\
\int_a^c q(t)\dt < \infty,\quad\int_c^x q(t)\dt \to\infty\quad(x\to b)
\end{gather*}
for some fixed but arbitrary point $c\in(a,b)$. In addition,
\begin{equation*}
\lim_{x\to a+}\frac{q(x)}{p(x)} = 0 = \lim_{x\to b-}\frac{p(x)}{q(x)}
\end{equation*}
\item $R:(a,b)\longrightarrow\MzC$ is an integrable matrix function.
\end{enumerate}

\begin{Lemma} \label{Limit} Let the coefficients of \eqref{CTHS} satisfy the conditions \cref{i}, \cref{ii} and $\alpha,\beta>0$. Then the associated differential expression \eqref{CTDE} is in the limit point case at both boundary points $a$, $b$. Moreover, for any $\lambda\in\C$ the differential system \eqref{CTDS} has a non-trivial solution $y(x)\not\equiv 0$ with the property
\begin{equation}
\int_a^b y(x)\A W(x)\,y(x)\dx < \infty \label{L2W}
\end{equation}
if and only if $\lim_{x\to a+} y(x) = \lim_{x\to b-} y(x) = 0$ holds.
\end{Lemma}

\begin{Proof}
We will study the behavior at the boundary point $a$, and for this we define
\begin{equation*}
T := \begin{pmatrix} 1 & \gamma+\lambda \\[1ex] 0 & -2\alpha \end{pmatrix}
\quad\mbox{and}\quad
S(x) := T^{-1}\left(q(x)\begin{pmatrix} \beta & \ms 0 \\[1ex] \gamma-\lambda & -\beta \end{pmatrix} - JR(x)\right)T
\end{equation*}
If we apply the transformation $y(x) = T\tilde y(x)$, then \eqref{CTDS} turns into the asymptotic diagonal system
\begin{equation*}
\tilde y'(x) = \left(p(x)\begin{pmatrix} \alpha & \ms 0 \\[1ex] 0 & -\alpha \end{pmatrix} 
+ S(x)\right)\tilde y(x)
\end{equation*}
where $S(x)$ is a Lebesgue-integrable matrix function on the interval $(a,c)$. By Levinson's Theorem on asymptotic integration, \eqref{CTDS} possesses a fundamental matrix of the form
\begin{equation*}
Y(x) = (T+o(1))\begin{pmatrix} \E^{-\alpha\phi(x)} & 0 \\[1ex] 0 & \E^{\alpha\phi(x)} \end{pmatrix},
\quad\mbox{where}\quad\phi(x) := \int_x^c p(t)\dt
\end{equation*}
From \cref{i} it follows that $\phi(x)\to\infty$ and hence $\E^{-\alpha\phi(x)} = o(1)\,\E^{\alpha\phi(x)}$ for $x\to a$. Moreover, \cref{i} implies $q(x)=o(1)\,p(x)$ for $x\to a$, and thus
\begin{equation*}
Y(x) = \E^{\alpha\phi(x)}\left(T\begin{pmatrix} 0 & 0 \\[1ex] 0 & 1 \end{pmatrix} + o(1)\right),\quad
W(x) = p(x)\left(\begin{pmatrix*} 0 & 0 \\[1ex] 0 & 1 \end{pmatrix*} + o(1)\right)
\end{equation*}
As the solutions of \eqref{CTDS} have the form $y(x)=Y(x) u$ with some vector $u=(u_1,u_2)\T\in\C^2$, we obtain
\begin{equation*}
y(x)\A W(x)\,y(x) = u\A Y(x)\A W(x)Y(x)u = p(x)\,\E^{2\alpha\phi(x)}(4\alpha^2 u_2^2 + o(1))
\end{equation*}
If $u_2\neq 0$, then $4\alpha^2 u_2^2\geq C>0$ holds in some neighborhood $(a,\delta)$ of $a$, and 
\begin{align*}
\int_\xi^\delta y(x)\A W(x)\,y(x)\dx & \geq \int_\xi^\delta C p(x)\,\E^{2\alpha\phi(x)}\dx 
= \frac{C}{2\alpha}\left(\E^{2\alpha\phi(\xi)}-\E^{2\alpha\phi(\delta)}\right)\to\infty
\end{align*}
as $\xi\to a$ implies that this solution does not lie left in $\LzW$. According to Weyl's alternative (see e.\,g. \cite[Theorem 5.6]{Weidmann:1987}) $\tau$ is in the limit point case at $a$. For \eqref{L2W} being valid, $u_2=0$ must be fulfilled, and the corresponding solution
\begin{equation*}
y(x) = (T+o(1))\begin{pmatrix} \E^{-\alpha\phi(x)} & 0 \\[1ex] 0 & \E^{\alpha\phi(x)}\end{pmatrix}\begin{pmatrix} u_1 \\[1ex] 0 \end{pmatrix}
= u_1\E^{-\alpha\phi(x)}(T+o(1))\,\hat{e}_1\quad\mbox{with}\quad 
\hat{e}_1 := \begin{pmatrix} 1 \\[1ex] 0 \end{pmatrix}
\end{equation*}
satisfies $\lim_{x\to a+}y(x) = 0$. Conversely, if  
\begin{equation*}
y(x) 
= (T+o(1))\begin{pmatrix} u_1\E^{-\alpha\phi(x)} \\[1ex] u_2\E^{\alpha\phi(x)} \end{pmatrix} 
= \E^{\alpha\phi(x)}\begin{pmatrix} (\gamma+\lambda)u_2 + o(1) \\[1ex] -2\alpha u_2 + o(1) \end{pmatrix}\to 0\quad\mbox{as}\quad x\to a
\end{equation*}
then $\E^{\alpha\phi(x)}\to\infty$ yields $u_2=0$, and $y(x) = u_1\E^{-\alpha\phi(x)}(T+o(1))\,\hat{e}_1$ gives 
\begin{equation*}
y(x)\A W(x)y(x) = u_1^2 q(x)\,\E^{-2\alpha\phi(x)}(1+o(1)) = p(x)\,\E^{-2\alpha\phi(x)}o(1)
\end{equation*}
Hence, there is a constant $C\geq 0$ such that $y(x)\A W(x) y(x)\leq C p(x)\,\E^{-2\alpha\phi(x)}$ on some interval $(a,\delta)$, and 
\begin{equation*}
\int_a^\delta y(x)\A W(x)y(x)\dx\leq\int_a^\delta C p(x)\,\E^{-2\alpha\phi(x)}\dx 
= \frac{C}{2\alpha}\,\E^{-2\alpha\phi(\delta)}
\end{equation*}
implies that $y(x)\A W(x)\,y(x)$ is integrable near $a$. By a similar reasoning we can prove that $\tau$ is in the limit point case at $b$ and that the integrability of $y(x)\A W(x)\,y(x)$ near $b$ is equivalent to $\lim_{x\to b-}y(x) = 0$.
\end{Proof}

\section{Linear $2\times 2$ systems with regular singular boundary points}
\label{sec:RegSingular}

After the general considerations in section \ref{sec:CompTriang}, we will now deal with linear $2\times 2$ systems having the special form
\begin{equation}
z'(x) = \left(\tfrac{1}{x}\,A + \tfrac{1}{1-x}\,B + C(x;u_1,\ldots,u_m)\right)z(x),\quad x\in (0,1) \label{GenRS}
\end{equation}
with regular-singular points at $x=0$ and $x=1$. We suppose that the non-singular part of the coefficient matrix is given by
\begin{equation*}
C(x;u_1,\ldots,u_m) = \sum_{k=1}^m u_k C_k(x)
\end{equation*}
where $C_k:[0,1]\longrightarrow\MzR$ for $k=1,\ldots,m$ are fixed matrix functions of class $\mathrm{C}^2$ and $u_1,\ldots,u_m\in\R$ are free parameters. Moreover, we require that $A,B\in\MzR$ satisfy condition \eqref{CTM}, and as stated at the beginning of section \ref{sec:CompTriang}, we may assume $\tr C(x)\equiv 0$ without any restriction. Now, according to Theorem \ref{SysCTF} there is a matrix $T\in\GLzR$, so that $z(x) = T y(x)$ transforms \eqref{GenRS} to the linear Hamiltonian system
\begin{equation}
J y'(x) = \left(
\frac{1}{x}\begin{pmatrix} 0 & \alpha \\[1ex] \alpha & \lambda+\gamma \end{pmatrix} + 
\frac{1}{1-x}\begin{pmatrix} \lambda-\gamma & \beta \\[1ex] \beta & 0 \end{pmatrix} + 
\sum_{k=1}^m u_k H_k(x)\right)
y(x) \label{CTRS}
\end{equation}
on the interval $(0,1)$. Here, $\alpha,\beta>0$ and $\gamma,\lambda\in\R$ are some real numbers, and $H_k:[0,1]\longrightarrow\MzR$ are twice continuously differentiable matrix functions satisfying $H_k(x)\A=H_k(x)$ on $[0,1]$  for $k=1,\ldots,m$. If we define
\begin{equation} \label{JW}
J := \begin{pmatrix} 0 & -1 \\[1ex] 1 & \ms 0 \end{pmatrix},\quad
W(x) := \begin{pmatrix} \frac{1}{1-x} & 0 \\[1ex] 0 & \frac{1}{x} \end{pmatrix},\quad
H_0(x) := \frac{1}{x}\begin{pmatrix} 0 & \alpha \\[1ex] \alpha & \gamma \end{pmatrix} + \frac{1}{1-x}\begin{pmatrix} -\gamma & \beta \\[1ex] \ms\beta & 0 \end{pmatrix}
\end{equation}
then the Hamiltonian system \eqref{CTRS} can be written as an eigenvalue problem $\tau y = \lambda y$ with the differential expression
\begin{equation*}
\tau y := W(x)^{-1}\left(J y'(x) - \left(H_0(x) + \sum_{k=1}^m u_k H_k(x)\right)y(x)\right)
\end{equation*}
We may consider $\tau=\tau(u_1,\ldots,u_m)$ to be an analytical perturbation of the differential expression
\begin{equation*}
\tau_0 y := W(x)^{-1}\big(J y'(x) - H_0(x)y(x)\big)
\end{equation*}
where $u_1=\ldots=u_m=0$. The coefficients $p(x)=\frac{1}{x}$ and $q(x)=\frac{1}{1-x}$ on $(0,1)$ satisfy the condition \cref{i} from the last section, and since the matrix function $R(x) := \sum_{k=1}^m u_k H_k(x)$ is continuous on $[0,1]$, it fulfills condition \cref{ii}. Therefore, $\tau_0$ is in the limit point case at $x=0$ and $x=1$ according to Lemma \ref{Limit}, and the minimal operator $\Tmin$ generated by $\tau_0$ possesses a unique self-adjoint extension $T_0 = \Tcls$. Hence there exists a self-adjoint operator $T_0$ with domain $D(T_0)\subset\LWz((0,1),\Cz)$, which is associated to the differential expression $\tau_0$ in a natural way. As $H_k(x)$ and $W(x)^{-1/2}$ for $k\in\{1,\ldots,m\}$ are continuous on $[0,1]$, we additionally obtain that the matrix functions $W(x)^{-1/2}H_k(x) W(x)^{-1/2}$ are bounded. This means that \eqref{CTRS} fulfills the assumptions \cref{A}, \cref{B} listed in section \ref{sec:HamiltonSys}. According to Theorem \ref{Family} we can now associate a holomorphic family of self-adjoint operators to the differential expressions $\tau$ by means of $T y = \tau y$, $y\in D(T_0)$. Finally, if we want to comply with condition \cref{C}, then we have to determine the (simple) eigenvalues of the unperturbed system. In the case $u_1=\ldots=u_m=0$ the differential equation $\tau_0 y=\lambda y$ corresponds to the system
\begin{equation}
y'(x) = \begin{pmatrix} \frac{\alpha}{x}+\frac{\beta}{1-x} & \frac{\gamma+\lambda}{x} \\[1ex] \frac{\gamma-\lambda}{1-x} & -\frac{\alpha}{x}-\frac{\beta}{1-x} \end{pmatrix}y(x),\quad
x\in(0,1) \label{HypGeo}
\end{equation}
We need to find values $\lambda\in\R$ for which \eqref{HypGeo} has a non-trivial solution $y(x)\neq 0$ with the property
\begin{equation}
\int_0^1 y(x)\A\begin{pmatrix}\frac{1}{1-x} & 0 \\[1ex] 0 & \frac{1}{x} \end{pmatrix}y(x)\dx<\infty \label{IntCon}
\end{equation}

\begin{Lemma} \label{Jacobi}
Let $\alpha,\beta>0$ and $\gamma\in\R$. The linear differential system \eqref{HypGeo} has a nontrivial solution $y(x)$ satisfying \eqref{IntCon} if and only if
\begin{equation*}
\lambda^2 = \gamma^2 + (2\alpha+1+n)(2\beta+1+n)
\end{equation*}
with some non-negative integer $n$. In this case the corresponding solutions are given by
\begin{equation*}
y(x) = C\,x^\alpha(1-x)^\beta\begin{pmatrix} 
(2\alpha+1+n)(1-x)P_n^{(2\alpha,2\beta+1)}(1-2x) \\[1ex] 
(\gamma-\lambda)\,x P_n^{(2\alpha+1,2\beta)}(1-2x) 
\end{pmatrix}
\end{equation*}
where $C\in\C$ is some constant, and $P_n^{(\alpha,\beta)}(z)$ denotes the Jacobi polynomial of degree $n$ with respect to the weight function $(1-z)^\alpha(1+z)^\beta$.
\end{Lemma}

\begin{Proof} By means of the transformation 
\begin{equation}
y(x) = x^{\alpha}(1-x)^{1+\beta}\begin{pmatrix} u(x) \\[0.5ex] v(x) \end{pmatrix} \label{Trafo}
\end{equation}
the linear system \eqref{HypGeo} turns to the pair of differential equations
\begin{align}
u'(x) & = \tfrac{1+2\beta}{1-x}\,u(x) + \tfrac{\gamma+\lambda}{x}\,v(x) \label{HypGeoA} \\
v'(x) & = \tfrac{\gamma-\lambda}{1-x}\,u(x) + \left(\tfrac{1}{1-x}-\tfrac{2\alpha}{x}\right)v(x)
\label{HypGeoB}
\end{align}
and the weighted square-integrability condition \eqref{IntCon} is equivalent to
\begin{equation}
\int_0^1 x^{2\alpha}(1-x)^{1+2\beta}u(x)^2\dx < \infty \quad\mbox{and}\quad
\int_0^1 x^{2\alpha-1}(1-x)^{2+2\beta}v(x)^2\dx < \infty \label{IntSep}
\end{equation}
We first consider the case $\gamma+\lambda=0$, where equation \eqref{HypGeoA} has the solution $u(x)=u_0(1-x)^{-1-2\beta}$ with some constant $u_0$. If $u_0\neq 0$, then the left integral in \eqref{IntSep} is not finite because of $\beta>0$. Hence $u(x)\equiv 0$, and \eqref{HypGeoB} yields $v(x)=v_0 x^{-2\alpha}(1-x)^{-1}$ with some constant $v_0$, where the second condition in \eqref{IntSep} and $\alpha>0$ yield $v_0=0$, so that for $\gamma+\lambda=0$ only $y\equiv 0$ satisfies condition \eqref{IntCon}. Since we are interested in nontrivial solutions, we may assume $\gamma+\lambda\neq 0$, and we can solve \eqref{HypGeoA} for $v(x)$:
\begin{equation} \label{Resolve}
v(x) = \frac{x\left(u'(x)-\tfrac{1+2\beta}{1-x}\,u(x)\right)}{\gamma+\lambda}
\end{equation}
After inserting this expression in \eqref{HypGeoB}, we receive for $u(x)$ the differential equation
\begin{equation*}
x(1-x)u''(x)+(1+2\alpha-(3+2\alpha+2\beta)x)u'(x)-(\gamma^2-\lambda^2+4\alpha\beta+2\alpha+2\beta+1)u(x) = 0
\end{equation*}
This is a hypergeometric differential equation
\begin{equation*}
x(1-x)u''(x) + (c-(a+b+1)x)u'(x) - ab\,u(x) = 0
\end{equation*}
with parameters
\begin{equation*}
a := 1+\alpha+\beta-\nu,\quad b := 1+\alpha+\beta+\nu,\quad c := 1+2\alpha > 0
\end{equation*}
where $\nu := \sqrt{(\alpha-\beta)^2-\gamma^2+\lambda^2}$. In addition to the hypergeometric function $u_1(x)=F(a,b;c;x)$ there is a second fundamental solution $u_2(x) = x^{-2\alpha}F(a-c+1,b-c+1;2-c;x)$ which behaves like $u_2(x)\sim x^{-2\alpha}$ as $x\to 0$. For the first entry in the corresponding solution $y(x)$ of \eqref{HypGeo} we get $x^{\alpha}(1-x)^{1+\beta}u_2(x)\sim x^{-\alpha}\to\infty$ as $x\to 0$, and this solution does not satisfy \eqref{IntCon} according to Lemma \ref{Limit} (alternatively one can use \eqref{Resolve} to prove that $v_2(x)=\left(-2\alpha/(\gamma+\lambda)+o(1)\right)x^{-2\alpha}$ as $x\to 0$, so that the second integral in \eqref{IntSep} is not finite). Furthermore, since $b>1$ and $c-a-b=-1-2\beta<-1$ hold, we obtain for $u_1(x)$ in case of $a\not\in-\N$ the asymptotic behavior
\begin{equation*}
u_1(x) = (1-x)^{-1-2\beta}\left(\frac{\Gamma(1+2\alpha)\,\Gamma(1+2\beta)}{\Gamma(a)\,\Gamma(b)}+o(1)\right)
\end{equation*}
as $x\to 1$ according to the linear transformation formulas (see e.\,g. \cite[Section 2.4.1]{MOS:1966}), and for this function the left integral in \eqref{IntSep} is not finite (it is not integrable at $x=1$). On the other hand, if $a=-n$ or equivalently $\nu=1+\alpha+\beta+n$ holds with some non-negative integer $n$, then $u_1(x) = F(a,b;c;x)$ reduces to a polynomial of degree $n$, which fulfills the first condition in \eqref{IntSep}. The constraint $a=-n$ provides the desired condition for the eigenvalue parameter:
\begin{equation*}
\lambda^2 = \gamma^2 + \nu^2-(\alpha-\beta)^2 = \gamma^2 + (2\alpha+n+1)(2\beta+n+1)
\end{equation*}
Additionally, by using $b = 1+\alpha+\beta+\nu=2\alpha+2\beta+2+n$, the solution $u_1(x)$ can be written as a constant multiple of a Jacobi polynomial (cf. \cite[15.4.6]{AS:1984}):
\begin{equation*}
u_1(x) = F(-n,2\alpha+1+2\beta+1+n;2\alpha+1;x) = \frac{n!}{(2\alpha+1)_n}P_n^{(2\alpha,2\beta+1)}(1-2x)
\end{equation*}
where $(a)_n$ denotes the Pochhammer symbol. We still have to calculate the corresponding function $v_1(x)$ by means of \eqref{Resolve}. Applying the differentiation formulas and Gauss's contiguous relations (see \cite[15.2.1 and 15.2.15]{AS:1984}) to the hypergeometric function $u_1(x)=F(a,b;c;x)$, we get 
\begin{align*}
(1-x)u_1'(x) 
& = (1-x)\tfrac{ab}{c}F(a+1,b+1;c+1;x) \\
& = \tfrac{b}{c}\big((c-b)F(a,b;c+1;x) - (c-a-b)F(a,b+1;c+1;x)\big) \\
& = \tfrac{b(c-b)}{c}F(a,b;c+1;x) - \tfrac{c-a-b}{c}\,b\,F(a,b+1;c+1;x)
\end{align*}
Moreover, $b F(a,b+1;c+1;x) = c F(a,b;c;x) - (c-b) F(a,b;c+1;x)$ (cf. \cite[15.2.24]{AS:1984}) implies
\begin{equation*}
(1-x)u_1'(x) = \tfrac{(c-a)(c-b)}{c}F(a,b;c+1;x) - (c-a-b)F(a,b;c;x)
\end{equation*}
where $c-a-b=-1-2\beta$ and $(c-a)(c-b)=(\alpha-\beta)^2-\nu^2=\gamma^2-\lambda^2$, so that
\begin{align*}
(1-x)u_1'(x) - (1+2\beta)u_1(x) = \tfrac{\gamma^2-\lambda^2}{c}F(a,b;c+1;x)
\end{align*}
and therefore
\begin{align*}
v_1(x) 
& = \frac{x\left((1-x)u_1'(x)-(1+2\beta)u_1(x)\right)}{(1-x)(\gamma+\lambda)}
  = \frac{x}{1-x}\frac{\gamma-\lambda}{2\alpha+1} F(a,b;c+1;x) \\
& = \frac{x}{1-x}\frac{\gamma-\lambda}{2\alpha+1}\frac{n!}{(2\alpha+2)_n}P_n^{(2\alpha+1,2\beta)}(1-2x)
  = \frac{x}{1-x}\frac{\gamma-\lambda}{2\alpha+1+n}\frac{n!}{(2\alpha+1)_n}P_n^{(2\alpha+1,2\beta)}(1-2x)
\end{align*}
If we multiply $u_1(x)$ and hence also $v_1(x)$ by $\frac{2\alpha+1+n}{n!}(2\alpha+1)_{n}$, then we get the functions
\begin{equation*}
u(x) = (2\alpha+1+n)P_n^{(2\alpha,2\beta+1)}(1-2x),\quad
v(x) = \frac{(\gamma-\lambda)x}{1-x}P_n^{(2\alpha+1,2\beta)}(1-2x)
\end{equation*}
which satisfy \eqref{HypGeoA}, \eqref{HypGeoB} and \eqref{IntSep}. Finally, the constant multiples of \eqref{Trafo} correspond to the solutions of \eqref{HypGeo}, \eqref{IntCon}.
\end{Proof}

So far we have transformed the parameter dependent differential system \eqref{CTRS} into an eigenvalue problem for a holomorphic family of self-adjoint differential operators $T=T(u_1,\ldots,u_m)$. Moreover, the eigenvalues of the unperturbed operator $T(0,\ldots,0)$ are given by Lemma \ref{Jacobi}. Each such eigenvalue is isolated with multiplicity one and can be continued to a simple eigenvalue $\lambda=\lambda(u_1,\ldots,u_m)$ of $T=T(u_1,\ldots,u_m)$, each of which is analytic in some neighborhood of $(0,\ldots,0)$. Now we will specify the dependency of the eigenvalues on the parameters $u_1,\ldots,u_m$ by means of a PDE.

\begin{Theorem} \label{PDE} 
Let $n$ be a non-negative integer and $\lambda:\Dom\longrightarrow\R$ be an analytic continuation of
\begin{equation}
\lambda_0 = \lambda(0,\ldots,0) = \pm\sqrt{\gamma^2 + (2\alpha+n+1)(2\beta+n+1)}
\label{EVRS}
\end{equation}
to some domain $\Dom\subset\R^m$ with $(0,\ldots,0)\in\Dom$, such that $\lambda=\lambda(u_1,\ldots,u_m)$ is a simple eigenvalue of the self-adjoint differential operator $T=T(u_1,\ldots,u_m)$ generated by the linear Hamiltonian system \eqref{CTRS}. If there exist a differentiable matrix function $G:[0,1]\times\C\times\Dom\longrightarrow\MznC$ and scalar functions $f_1,\ldots,f_m,g:\C\times\Dom\longrightarrow\C$ such that $H(x;u_1\ldots,u_m) = H_0(x)+\sum_{k=1}^m u_k H_k(x)$, $G=G(x;\lambda;u_1,\ldots,u_m)$ and $f_k=f_k(\lambda;u_1,\ldots,u_m)$, $g=g(\lambda;u_1,\ldots,u_m)$ fulfill the deformation equation
\begin{equation*}
\pD{G}{x} + (\lambda W + H)JG - GJ(\lambda W + H) = \sum_{k=1}^{m} f_k H_k + gW 
\end{equation*}
then the eigenvalues $\lambda=\lambda(u_1,\ldots,u_m)$ satisfy on $\Dom$ the quasilinear partial differential equation
\begin{equation*}
\sum_{k=1}^{m} f_k(\lambda;u_1,\ldots,u_m)\pD{\lambda}{u_k} = g(\lambda;u_1,\ldots,u_m)
\end{equation*}
\end{Theorem}

\begin{Proof} The operators $T(u_1,\ldots,u_m)$, the eigenvalue $\lambda_0$ of $T(0,\ldots,0)$, and the functions $G$, $f_k$, $g$ meet the conditions \cref{A} -- \cref{D} in section \ref{sec:HamiltonSys}. Further, an eigenfunction $y(x)$ of $T(u_1,\ldots,u_m)$ associated with $\lambda(u_1,\ldots,u_m)$ satisfies
\begin{equation*}
\int_0^1 y(x)\A W(x)\,y(x)\dx < \infty
\end{equation*}
and hence Lemma \ref{Limit} implies $\lim_{x\to 0+}y(x) = \lim_{x\to 1-}y(x) = 0$. Since $G(x)$ is bounded on $[0,1]$, we obtain
\begin{equation*}
\lim_{x\to 0+}y(x)\A G(x;\lambda;u_1,\ldots,u_m)y(x) 
= \lim_{x\to 1-}y(x)\A G(x;\lambda;u_1,\ldots,u_m)y(x) = 0
\end{equation*}
From Theorem \ref{Family} it follows that $\lambda=\lambda(u_1,\ldots,u_m)$ solves the PDE \eqref{PDE0} on $\Dom$.
\end{Proof}

\subsection{Linear $2\times 2$ systems of the type $z'(x) = \big(\frac{1}{x}A+\frac{1}{1-x}B+C\big)z(x)$ on $(0,1)$}\label{sec:RegSingC}\subsep

As a first special case of \eqref{GenRS} we consider linear differential systems having the form
\begin{equation}
z'(x) = \left(\tfrac{1}{x}\,A + \tfrac{1}{1-x}\,B + C\right)z(x),\quad x\in (0,1) \label{ConRS}
\end{equation}
with regular singular points at $x=0$ and $x=1$, where $C\in\MzR$ is a constant matrix. If $A$, $B$ match condition \eqref{CTM} and $\tr C=0$ is fulfilled, then we can transform \eqref{ConRS} into the complementary triangular form
\begin{equation}
J y'(x) = \left(
\frac{1}{x}\begin{pmatrix} 0 & \alpha \\[1ex] \alpha & \lambda+\gamma \end{pmatrix} + 
\frac{1}{1-x}\begin{pmatrix} \lambda-\gamma & \beta \\[1ex] \beta & 0 \end{pmatrix} + 
\begin{pmatrix} u_1 & u_2 \\[1ex] u_2 & u_3 \end{pmatrix}\right)y(x)
\label{GCHE}
\end{equation}
This conversion process is outlined in section \ref{sec:CompTriang}. If we assume $\alpha,\beta>0$ and $\gamma\in\R$ to be fixed, and if $\lambda\in\R$ becomes the eigenvalue parameter, then there are three  values $u_1,u_2,u_3\in\R$ left, which will be considered as free parameters. By introducing the matrix functions
\begin{equation*}
W(x) := \begin{pmatrix} \frac{1}{1-x} & 0 \\[1ex] 0 & \frac{1}{x} \end{pmatrix},\quad
H(x) := \frac{1}{x}\begin{pmatrix} 0 & \alpha \\[1ex] \alpha & \gamma \end{pmatrix} + 
\frac{1}{1-x}\begin{pmatrix} -\gamma & \beta \\[1ex] \ms\beta & 0 \end{pmatrix} + 
\begin{pmatrix} u_1 & u_2 \\[1ex] u_2 & u_3 \end{pmatrix}
\end{equation*}
the system \eqref{ConRS} can be written in the form $\tau y = \lambda y$ with $\tau y := W^{-1}(J y'-H y)$, where in case of $u_1=u_2=u_3=0$ the associated self-adjoint operator has the eigenvalues \eqref{EVRS}.

\begin{Theorem} \label{EigRS} 
If \eqref{CTM} holds, then we can convert \eqref{ConRS} to a linear Hamiltonian system
\begin{equation*}
J y'(x) = \left(
\frac{1}{x}\begin{pmatrix} 0 & \alpha \\[1ex] \alpha & \lambda+\gamma \end{pmatrix} + 
\frac{1}{1-x}\begin{pmatrix} \lambda-\gamma & \beta \\[1ex] \beta & 0 \end{pmatrix} + 
\begin{pmatrix} u_1 & u_2 \\[1ex] u_2 & u_3 \end{pmatrix}\right)y(x)
\end{equation*}
with $\alpha,\beta > 0$ and $\gamma\in\R$, which can also be written as an eigenvalue problem $\tau y = \lambda y$. The self-adjoint differential operators $T=T(u_1,u_2,u_3)$ generated by $\tau$ in $\LWz((0,1),\Cz)$ have simple eigenvalues $\lambda=\lambda(u_1, u_2,u_3)$, which depend analytically on the parameters $u_1,u_2,u_3$ in some domain $\Dom\subset\R^3$ with $(0,0,0)\in\Dom$. For $u_1=u_2=u_3=0$ these eigenvalues are given by
\begin{equation*}
\lambda = \lambda(0,0,0) = \pm\sqrt{\gamma^2 + (2\alpha+n+1)(2\beta+n+1)}
\end{equation*}
with some non-negative integer $n$. Moreover, the eigenvalues $\lambda=\lambda(u_1,u_2,u_3)$ satisfy the partial differential equation
\begin{equation}\begin{split}
& \big((2\lambda+u_1)((\alpha-\beta)u_1+(\lambda-\gamma)u_2) + u_1(\lambda-\alpha u_1+\alpha u_3)\big)\pD{\lambda}{u_1} \\
& \quad {} + \lambda((\gamma+\lambda)(u_1-u_3)+(2\lambda+u_1)u_3 + u_2)\pD{\lambda}{u_2} \\
& \quad {} + \big((2\lambda+u_3)((\beta-\alpha)u_3+(\lambda+\gamma)u_2) + u_3(\lambda+\beta u_1-\beta u_3)\big)\pD{\lambda}{u_3} \\
& = \beta(\lambda+\gamma)u_1 + \alpha(\lambda-\gamma)u_3 + (\gamma^2-\lambda^2)u_2 \label{PDERS}
\end{split}\end{equation}
\end{Theorem}

\begin{Proof}
The non-singular part of the coefficient matrix $H$ is a linear combination of the symmetric matrices
\begin{equation*}
H_1 := \begin{pmatrix} 1 & 0 \\[1ex] 0 & 0 \end{pmatrix},\quad
H_2 := \begin{pmatrix} 0 & 1 \\[1ex] 1 & 0 \end{pmatrix},\quad
H_3 := \begin{pmatrix} 0 & 0 \\[1ex] 0 & 1 \end{pmatrix} 
\end{equation*}
In particular $H$ has the form \eqref{LinearH} with $m=3$. Further we define the scalar functions
\begin{align*}
f_1(\lambda;u_1,u_2,u_3) & := 
(2\lambda+u_1)((\alpha-\beta)u_1+(\lambda-\gamma)u_2) + u_1(\lambda-\alpha u_1+\alpha u_3) \\
f_2(\lambda;u_1,u_2,u_3) & := \lambda((\gamma+\lambda)(u_1-u_3) + (2\lambda+u_1)u_3 + u_2) \\
f_3(\lambda;u_1,u_2,u_3) & := 
(2\lambda+u_3)((\beta-\alpha)u_3+(\lambda+\gamma)u_2) + u_3(\lambda+\beta u_1-\beta u_3) \\
  g(\lambda;u_1,u_2,u_3) & :=
\beta(\lambda+\gamma)u_1 + \alpha(\lambda-\gamma)u_3 + (\gamma^2-\lambda^2)u_2
\end{align*}
and for $(x;\lambda;u_1,u_2,u_3)\in[0,1]\times\C\times\R^3$ the differentiable matrix function
\begin{equation*}
G(x;\lambda;u_1,u_2,u_3) := \begin{pmatrix} 
\lambda u_1 x & \lambda u_2 x + \frac{\beta u_1 - \alpha u_3 + (\gamma-\lambda)u_2}{2} \\[1ex] 
\lambda u_2 x + \frac{\beta u_1 - \alpha u_3 + (\gamma-\lambda)u_2}{2} & \lambda u_3(x-1) \end{pmatrix} 
\end{equation*}
A direct but cumbersome calculation proves that these functions satisfy the deformation equation \eqref{DefLin}, and hence the eigenvalues solve the partial differential equation $f_1\pD{\lambda}{u_1} + f_2\pD{\lambda}{u_2} + f_3\pD{\lambda}{u_3} = g$ due to Theorem \ref{PDE}.
\end{Proof}

The special case $\alpha=\beta$, $\gamma=0$ not only simplifies the coefficients in the eigenvalue PDE: We can set $u_3=u_1$ and thus decrease the number of free parameters by one. For such a linear system
\begin{equation}
J y'(x) = \left(
\frac{1}{x}\begin{pmatrix} 0 & \alpha \\[1ex] \alpha & \lambda \end{pmatrix} + 
\frac{1}{1-x}\begin{pmatrix} \lambda & \alpha \\[1ex] \alpha & 0 \end{pmatrix} + 
\begin{pmatrix} u_1 & u_2 \\[1ex] u_2 & u_1 \end{pmatrix}\right)y(x) \label{CPDS}
\end{equation}
with $\alpha>0$ and two parameters $u_1,u_2\in\R$ we get the following result for the eigenvalues:

\begin{Theorem}
In the case $u_1=u_2=0$ the eigenvalues associated to \eqref{CPDS} are given by
\begin{equation*}
\lambda = \lambda(0,0) = \pm(2\alpha+n+1)\quad\mbox{for}\quad n=0,1,2,3,\ldots
\end{equation*}
They can be analytically continued to a domain $\Dom\subset\R^2$ with $(0,0)\in\Dom$ so that $\lambda=\lambda(u_1,u_2)$ are simple eigenvalues of \eqref{CPDS} for all $(u_1,u_2)\in\Dom$. Moreover, these eigenvalues satisfy the partial differential equation
\begin{equation}
(u_1+u_1 u_2 + 2\lambda u_2)\pD{\lambda}{u_1} + (u_2+u_1^2 + 2\lambda u_1)\pD{\lambda}{u_2} 
= 2\alpha u_1 - \lambda u_2 \label{CPEV}
\end{equation}
\end{Theorem}

\begin{Proof}
From Lemma \ref{Jacobi} with $\beta=\alpha$ and $\gamma=0$ we get the eigenvalues for $u_1=u_2=0$. The coefficient matrix $H$ has the form \eqref{LinearH} with $m=2$ and
\begin{equation*}
H_1 := \begin{pmatrix} 1 & 0 \\[1ex] 0 & 1 \end{pmatrix},\quad
H_2 := \begin{pmatrix} 0 & 1 \\[1ex] 1 & 0 \end{pmatrix}
\end{equation*}
A somewhat tedious calculation confirms that the functions
\begin{gather*}
f_1 := u_1 + u_1 u_2 + 2\lambda u_2, \quad
f_2 := u_2 + u_1^2 + 2\lambda u_1, \quad
  g := 2\alpha u_1-\lambda u_2 \\
G = G(x;\lambda;u_1,u_2) 
    := \begin{pmatrix} u_1 x & u_2(x - \frac{1}{2}) \\[1ex] u_2(x - \frac{1}{2}) & u_1(x-1) \end{pmatrix} 
\end{gather*}
satisfy the deformation equation \eqref{DefLin}, and finally Theorem \ref{PDE} yields the partial differential equation $f_1\pD{\lambda}{u_1} + f_2\pD{\lambda}{u_2} = g$ for the eigenvalues, which corresponds to the PDE \eqref{CPEV}.
\end{Proof}

The differential system \eqref{CPDS} is a generalization of the Chandrasekhar-Page angular equation \eqref{CPJH}, and therefore it should be possible to derive the partial differential equation \eqref{BSWE} from \eqref{CPEV}. For this reason we first have to rearrange the coefficient matrix of \eqref{CPJH}. In the form
\begin{equation*}
J y' = \left(\frac{1}{x}
\begin{pmatrix} 0 & -\frac{2\kappa+1}{4} \\[1ex] -\frac{2\kappa+1}{4} & \mu-\Lambda \end{pmatrix} 
+ \frac{1}{1-x}\begin{pmatrix} \mu-\Lambda & -\frac{2\kappa+1}{4} \\[1ex] -\frac{2\kappa+1}{4} & 0 \end{pmatrix} + 
\begin{pmatrix} -2\mu & -2\nu \\[1ex] -2\nu & -2\mu \end{pmatrix}\right)y
\end{equation*}
the CPAE complies with the linear Hamiltonian system \eqref{CPDS} if we set
\begin{equation*}
\alpha = -\tfrac{1}{4}(2\kappa+1),\quad u_1=-2\mu,\quad u_2 = -2\nu,\quad
\lambda = \mu-\Lambda = -\tfrac{1}{2} u_1-\Lambda
\end{equation*}
Because of $\frac{\partial\lambda}{\partial u_1}=-\frac{1}{2}+\frac{1}{2}\frac{\partial\Lambda}{\partial\mu}$ and $\frac{\partial\lambda}{\partial u_2} = \frac{1}{2}\frac{\partial\Lambda}{\partial\nu}$ the PDE \eqref{CPEV} turns into \eqref{BSWE}:
\begin{align*}
0 & 
= 2\alpha u_1-\lambda u_2 - (u_1+u_1 u_2+2u_2\lambda)\pD{\lambda}{u_1} - (u_2+u_1^2+2u_1\lambda)\pD{\lambda}{u_2} \\
& = (2\kappa+1)\mu + 2\nu(\mu-\Lambda) + (\mu-2\nu\Lambda)\left(\frac{\partial\Lambda}{\partial\mu}-1\right) + (\nu-2\mu\Lambda)\frac{\partial\Lambda}{\partial\nu} \\
& = 2\kappa\mu + 2\mu\nu + (\mu-2\nu\Lambda)\frac{\partial\Lambda}{\partial\mu} + (\nu-2\mu\Lambda)\frac{\partial\Lambda}{\partial\nu}
\end{align*}

The linear system \eqref{GCHE} may also be regarded as a generalization of the confluent Heun equation (CHE), as the following considerations show. Multiplying \eqref{GCHE} by $-J=J^{-1}$ from the left gives
\begin{equation*}
y'(x) = \left(
\frac{1}{x}\begin{pmatrix} \alpha & \gamma+\lambda \\[1ex] 0 & -\alpha \end{pmatrix} 
+ \frac{1}{1-x}\begin{pmatrix} \beta & \ms 0 \\[1ex] \gamma-\lambda & -\beta \end{pmatrix} 
+ \begin{pmatrix*}[r] u_2 & u_3 \\[1ex] -u_1 & -u_2 \end{pmatrix*}
\right)y(x) 
\end{equation*}
By means of the transformation
\begin{equation*}
y(x) = x^{\alpha}(1-x)^{-\beta}\E^{u_2 x}\begin{pmatrix} w(x) \\[1ex] \tilde w(x) \end{pmatrix}
\end{equation*}
this differential system is equivalent to
\begin{equation*}
\begin{pmatrix} w'(x) \\[1ex] \tilde w'(x) \end{pmatrix} = \begin{pmatrix} 
0 & \frac{\gamma+\lambda}{x} + u_3 \\[1ex] 
\frac{\gamma-\lambda}{1-x} - u_1 & -\frac{2\alpha}{x} - \frac{2\beta}{1-x} - 2u_2 \end{pmatrix}
\begin{pmatrix} w(x) \\[1ex] \tilde w(x) \end{pmatrix}
\end{equation*}
Solving the upper equation $w'(x) = \left(\frac{\gamma+\lambda}{x} + u_3\right)\tilde w(x)$ for $\tilde w(x)$ and computing $\tilde w'(x)$, we obtain the second order ODE $w''(x) + f(x)w'(x) + g(x)w(x) = 0$, where
\begin{align*}
f(x) & = \frac{2\alpha+1}{x} - \frac{2\beta}{x-1} - \frac{u_3}{\gamma+\lambda + x u_3} + 2u_2 \\
g(x) & = \frac{(\gamma+\lambda)(\lambda-\gamma+u_1)}{x} + \frac{(\gamma-\lambda)(\lambda+\gamma+u_3)}{x-1} + u_1 u_3
\end{align*}
For the special case $u_3=0$ we get the differential equation
\begin{equation*}
w''(x) 
+ \left(\frac{2\alpha+1}{x} + \frac{-2\beta}{x-1} + 2 u_2 \right)w'(x) 
+ \frac{(\gamma+\lambda)u_1 x + (\gamma+\lambda)(\gamma-\lambda-u_1)}{x(x-1)}w(x) = 0
\end{equation*}
which coincides with the non-symmetrical canonical form of the CHE (see \cite[part B, section 1.2]{Ronveaux:1995})
\begin{equation*}
w''(x) 
+ \left(\frac{\gamma}{x} + \frac{\delta}{x-1} + \varepsilon\right)w'(x) 
+ \frac{ax-q}{x(x-1)}\,w(x) = 0
\end{equation*}

\subsection{Linear $2\times 2$ systems $z'(x) = \big(\frac{1}{x}A+\frac{1}{1-x}B+\Polynom\big)z(x)$ on $(0,1)$}\subsep

We now turn to more general systems of the form
\begin{equation}
z'(x) = \left(\frac{1}{x}\,A + \frac{1}{1-x}\,B + \sum_{k=0}^m x^k C_k\right)z(x),\quad x\in (0,1) \label{PolRS}
\end{equation}
with regular singular boundary points, where we replace the constant matrix $C$ in section \ref{sec:RegSingC} by a polynomial of degree $m$. Again, we assume that the matrices $A,B\in\MzR$ satisfy \eqref{CTM}, and we may also suppose without restriction that $\tr C_k=0$ holds for $k=0,\ldots,m$. With these conditions, \eqref{PolRS} can be transformed to the complementary triangular form
\begin{gather*}
J y'(x) = (\lambda W(x) + H(x))y(x),\quad\mbox{where}\quad
W(x) := \begin{pmatrix} \frac{1}{1-x} & 0 \\[1ex] 0 & \frac{1}{x} \end{pmatrix}\quad\mbox{and} \\
H(x) := \frac{1}{x}\begin{pmatrix} 0 & \alpha \\[1ex] \alpha & \gamma \end{pmatrix} + 
\frac{1}{1-x}\begin{pmatrix} -\gamma & \beta \\[1ex] \ms\beta & 0 \end{pmatrix} + 
\sum_{k=0}^m x^k \begin{pmatrix} u_{3k+1} & u_{3k+2} \\[1ex] u_{3k+2} & u_{3k+3} \end{pmatrix}
\end{gather*}
Here $\alpha,\beta>0$ and $\gamma\in\R$ are fixed, and $\lambda\in\R$ becomes the eigenvalue parameter. In addition, this system contains $3m+3$ entries $u_1,u_2,\ldots,u_{3m+3}$, which are considered to be real parameters. We can write the system in the form $\tau y = \lambda y$ with $\tau y := W^{-1}(J y'- H y)$, and in case of $u_k=0$ for $k=1,2,\ldots,3m+3$ the eigenvalues of the associated self-adjoint differential operator are given by \eqref{EVRS}. If we define the functions
\begin{align*}
f_1 & := -Q u_1-\lambda((2\beta-2\alpha-1)u_{3m+1}+2(\gamma-\lambda)u_{3m+2}) \\
f_2 & := \lambda((\gamma+\lambda)u_{3m+1}+(u_1+\lambda-\gamma)u_{3m+3}+u_{3m+2}) \\
f_3 & :=  Q u_3+\lambda((2\beta-2\alpha+1+2 u_2)u_{3m+3}+2(\gamma+\lambda)u_{3m+2}) \\
  g & := \beta(\gamma+\lambda)u_{3m+1}-\alpha(\gamma-\lambda)u_{3m+3}+(\gamma^2-\lambda^2)u_{3m+2}
\end{align*}
with $Q := \beta u_{3m+1}-\alpha u_{3m+3}+(\gamma-\lambda) u_{3m+2}$ and
\begin{align*}
f_{3k+1} & := -Q u_{3k+1}+2\lambda(u_{3k-1} u_{3m+1}-u_{3k-2} u_{3m+2}) \\
f_{3k+2} & := (u_{3k} u_{3m+1}+(u_{3k+1}-u_{3k-2}) u_{3m+3})\lambda \\
f_{3k+3} & :=  Q u_{3k+3}+2\lambda(u_{3k} u_{3m+2}+(u_{3k+2}-u_{3k-1}) u_{3m+3})
\end{align*}
for $k=1,\ldots,m$, then the matrix function
\begin{equation*}
G(x;\lambda;u_1,u_2,\ldots,u_{3m+3}) := \begin{pmatrix} 
\lambda u_{3m+1} x & \lambda u_{3m+2} x + \frac{1}{2} Q \\[1ex]
\lambda u_{3m+2} x + \frac{1}{2} Q & \lambda u_{3m+3}(x-1)
\end{pmatrix}
\end{equation*}
solves the deformation equation
\begin{equation*}
\pD{G}{x} + (\lambda W + H)JG - GJ(\lambda W + H) = \sum_{k=1}^{3m+3} f_k\pD{H}{u_k} + gW
\end{equation*}
as a direct (but lengthy) calculation shows. Thus the simple eigenvalues $\lambda=\lambda(u_1,u_2,\ldots,u_{3m+3})$ of the self-adjoint operator associated to the differential expression $\tau=\tau(u_1,u_2,\ldots,u_{3m+3})$ satisfy the partial differential equation
\begin{equation*}
\sum_{k=1}^{3m+3} f_k(\lambda;u_1,u_2,\ldots,u_{3m+3})\pD{\lambda}{u_k} = g(\lambda;u_1,u_2,\ldots,u_{3m+3})
\end{equation*}

\subsection{Linear $2\times 2$ systems $z'(x) = \big(\frac{1}{x}A+\frac{1}{1-x}B+C+x(1-x)D\big)z(x)$ on $(0,1)$}\subsep

A system like this is a generalization of \eqref{ConRS} and also a special case of \eqref{PolRS} with a polynomial of degree $2$. By assuming \eqref{CTM} and $\tr C = \tr D = 0$, it can be converted to  complementary triangular form, and then we can write it as a linear Hamiltonian system
\begin{equation*}
J y'(x) = \left(
\frac{1}{x}\begin{pmatrix} 0 & \alpha \\[1ex] \alpha & \lambda+\gamma \end{pmatrix} + 
\frac{1}{1-x}\begin{pmatrix} \lambda-\gamma & \beta \\[1ex] \beta & 0 \end{pmatrix} + 
H_1 + x(1-x)H_2\right)y(x)
\end{equation*}
with symmetrical matrices $H_1$ and $H_2$, which have six entries in total. To keep it simple, we restrict our considerations to the case $\beta=\alpha$ and $\gamma=0$, so that
\begin{equation}
J y'(x) = \left(
\frac{1}{x}\begin{pmatrix} 0 & \alpha \\[1ex] \alpha & \lambda \end{pmatrix} + 
\frac{1}{1-x}\begin{pmatrix} \lambda & \alpha \\[1ex] \alpha & 0 \end{pmatrix}
+ \begin{pmatrix} u_1 & u_2 \\[1ex] u_2 & u_1 \end{pmatrix}
 + x(1-x)\begin{pmatrix} u_3 & u_4 \\[1ex] u_4 & u_3 \end{pmatrix}\right)y(x) \label{SymRS}
\end{equation}
has, in addition to $\alpha>0$ (fixed) and $\lambda$ (eigenvalue parameter), four entries $u_1$ to $u_4$ in the coefficient matrix, which are supposed to be real parameters. This linear Hamiltonian system has the form $J y' = (\lambda W(x) + H(x))y$, where $J$, $W(x)$ are given by \eqref{JW}, and
\begin{equation*}
H(x) := \frac{1}{x}\begin{pmatrix} 0 & \alpha \\[1ex] \alpha & 0 \end{pmatrix}
+ \frac{1}{1-x}\begin{pmatrix} 0 & \alpha \\[1ex] \alpha & 0 \end{pmatrix}
+ \begin{pmatrix} u_1 & u_2 \\[1ex] u_2 & u_1 \end{pmatrix}
+ x(1-x)\begin{pmatrix} u_3 & u_4 \\[1ex] u_4 & u_3 \end{pmatrix}
\end{equation*}
We have to solve the deformation equation, and for this reason we introduce the following scalar functions depending on $(u_1,u_2,u_3,u_4)$ and $\lambda$:
\begin{align*}
f_1 := & M(2\lambda u_4 + u_2 u_3 + u_3) + N\big(4\lambda u_2 - \lambda u_4 + 2 u_1 u_2 + 2 u_1 - u_3\big) \\
f_2 := & M(2\lambda u_3 + u_1 u_3 + u_4) + N\big(4\lambda u_1 - \lambda u_3 + 2 u_1^2 + 2 u_2 - u_4\big) \\
f_3 := & M u_3 u_4 + N(4\lambda u_4 + 2 u_1 u_4 + 6 u_3) \\
f_4 := & M u_3^2 + N(4\lambda u_3 + 2 u_1 u_3 + 6 u_4),\quad
  g := -M^2 + 2 N (2\alpha u_1 - \lambda u_2)
\end{align*}
where $M := \lambda u_4 - 2\alpha u_3$ and $N := u_2 u_3 - u_1 u_4$. If $H_0(x)$ denotes $H(x)$ with $u_1=u_2=u_3=u_4=0$, and if we define
\begin{equation*}
H_1 := \begin{pmatrix} 1 & 0 \\[1ex] 0 & 1 \end{pmatrix},\quad
H_2 := \begin{pmatrix} 0 & 1 \\[1ex] 1 & 0 \end{pmatrix},\quad
H_3(x) := x(1-x)H_1,\quad H_4(x) := x(1-x)H_2
\end{equation*}
then the coefficient matrix $H(x) := H_0(x) + u_1 H_1 + u_2 H_2 + u_3 H_3(x) + u_4 H_4(x)$ and the matrix function
\begin{equation*}\begin{split}
G(x;\lambda;u_1,u_2,u_3,u_4) := & \tfrac{1}{2} M
\begin{pmatrix} 2u_3 x  & u_4(2x-1) \\[1ex] u_4(2x-1) & 2u_3(x-1) \end{pmatrix} + {}
\\ & {} + N\left(\begin{pmatrix} 2 u_1 x & u_2(2x-1) \\[1ex] u_2(2x-1) & 2u_1(x-1) \end{pmatrix}
+ x(1-x)(2x-1)\begin{pmatrix} u_3 & u_4 \\[1ex] u_4 & u_3 \end{pmatrix}\right) \end{split}
\end{equation*}
satisfy the deformation equation \eqref{DefLin}, so that the eigenvalues $\lambda=\lambda(u_1,u_2,u_3,u_4)$ of the associated differential operator are solutions of the PDE 
\begin{equation*}
f_1\pD{\lambda}{u_1} + f_2\pD{\lambda}{u_2} + f_3\pD{\lambda}{u_3} + f_4\pD{\lambda}{u_4} = g
\end{equation*}
It is worth mentioning that by means of the transformation
\begin{equation*}
S(\theta) = \left(\begin{array}{cc} \sqrt{\tan\frac{\theta}{2}} & 0 \\[1ex]
0 & \sqrt{\cot\frac{\theta}{2}} \end{array}\right)y(\sin^2\tfrac{\theta}{2}),
\quad\theta\in(0,\pi)
\end{equation*}
the linear Hamiltonian system \eqref{SymRS} is on the interval $(0,\pi)$ equivalent to
\begin{equation*}
\begin{pmatrix*}[r] 0 & 1 \\[1ex] -1 & 0 \end{pmatrix*} S'(\theta) + \begin{pmatrix}
\frac{u_1}{2}\cos\theta + \frac{u_3}{8}(1+\cos\theta)\sin^2\theta &
\frac{4\alpha+1}{2\sin\theta} + \frac{u_2}{2}\sin\theta + \frac{u_4}{8}\sin^3\theta \\[1ex] 
\frac{4\alpha+1}{2\sin\theta} + \frac{u_2}{2}\sin\theta + \frac{u_4}{8}\sin^3\theta &
-\frac{u_1}{2}\cos\theta + \frac{u_3}{8}(1-\cos\theta)\sin^2\theta \end{pmatrix}S(\theta)
= \Lambda S(\theta)
\end{equation*}
where the eigenvalue parameter has been renamed to $\Lambda := -\frac{1}{2}u_1-\lambda$. This differential system is yet another generalization of the Chandrasekhar-Page angular equation \eqref{CPAE}, which not only contains the values $\alpha = -(2\kappa+1)/4$ and $u_1=-2\mu$, $u_2 = -2\nu$, but also two additional parameters $u_3$, $u_4$ as well as higher powers of trigonometric functions.

\section{Some more systems and associated PDEs for the eigenvalues}
\label{sec:NonRegular}

In this section we study a few more linear $2\times 2$ systems of the type \eqref{GDS} and at last also a $4\times 4$ system. Here, we no longer care about the conditions \cref{a} -- \cref{d} or \cref{A} -- \cref{D}. We will merely focus on the conversion to the Hamiltonian form and on finding a solution of the corresponding deformation equation, so that we are able to derive a partial differential equation for the eigenvalues.

\subsection{Linear $2\times 2$ systems of the type $z'(x) = \big(\frac{1}{x}A+\frac{1}{1-x}B+\frac{1}{x-t}C\big)z(x)$}\subsep

Such a system has, apart from $x=0$ and $x=1$, another regular singular point at $x=t$. Here we restrict our considerations to the interval $(0,1)$, and we assume $t\not\in[0,1]$. If the coefficient matrices $A$, $B$ fulfill condition \eqref{CTM}, then we can transform the system to the complementary triangular form
\begin{equation*}
y'(x) = \left(
\frac{1}{x}\begin{pmatrix} \alpha & \gamma+\lambda \\[1ex] 0 & -\alpha \end{pmatrix} + 
\frac{1}{1-x}\begin{pmatrix} \beta & \ms 0 \\[1ex] \gamma-\lambda & -\beta \end{pmatrix} + 
\frac{1}{x-t}\begin{pmatrix*}[r] u_2 & u_3 \\[1ex] -u_1 & -u_2 \end{pmatrix*}\right)y(x)
\end{equation*}
For convenience, we will only deal with the special case $\beta=\alpha$ and $\gamma=0$. Moreover, we consider $u_1, u_2, u_3$ as well as $u_4 = t$ to be real parameters. Multiplying this system from the left by $J$ gives
\begin{equation}
J y'(x) = \left(
\frac{1}{x}\begin{pmatrix} 0 & \alpha \\[1ex] \alpha & \lambda \end{pmatrix} + 
\frac{1}{1-x}\begin{pmatrix} \lambda & \alpha \\[1ex] \alpha & 0 \end{pmatrix} + 
\frac{1}{x-u_4}\begin{pmatrix} u_1 & u_2 \\[1ex] u_2 & u_3 \end{pmatrix}\right)y(x),
\quad x\in(0,1) \label{AddRS}
\end{equation}
The system is now written in Hamiltonian form $J y' = \left(\lambda W(x) + H(x)\right)y$, where $J$, $W(x)$ are given by \eqref{JW}, and
\begin{equation*}
H(x;u_1,u_2,u_3,u_4) := \frac{1}{x}\begin{pmatrix} 0 & \alpha \\[1ex] \alpha & 0 \end{pmatrix} + 
\frac{1}{1-x}\begin{pmatrix} 0 & \alpha \\[1ex] \alpha & 0 \end{pmatrix} + 
\frac{1}{x-u_4}\begin{pmatrix} u_1 & u_2 \\[1ex] u_2 & u_3 \end{pmatrix}
\end{equation*}
Further we define the five scalar functions
\begin{align*}
f_1 & := \alpha u_1 (u_3-u_1+2\lambda) + \lambda u_2 (u_1-2\lambda u_4) \\
f_2 & := \lambda u_1 (u_3+\lambda) - \lambda^2 u_4 (u_1+u_3) \\
f_3 & := \alpha u_3 (u_1-u_3-2\lambda) + \lambda u_2 (u_3-2\lambda u_4+2\lambda) \\
f_4 & := \lambda u_4 (u_4-1),\quad g := \alpha\lambda(u_1+u_3)-\lambda^2 u_2
\end{align*}
and for $(u_1,u_2,u_3,u_4)\in\R^3\times\R\setminus[0,1]$ the differentiable matrix function
\begin{equation*}
G(x;\lambda;u_1,u_2,u_3,u_4) := \begin{pmatrix} 
  \frac{\lambda u_1(1-u_4)x}{x-u_4}
& \frac{\alpha(u_1-u_3)+\lambda u_2}{2} + \frac{\lambda u_2 u_4 (1-x)}{x-u_4} \\[1ex] 
  \frac{\alpha(u_1-u_3)+\lambda u_2}{2} + \frac{\lambda u_2 u_4 (1-x)}{x-u_4} 
& \frac{\lambda u_3 u_4(1-x)}{x-u_4} \end{pmatrix} 
\end{equation*}
These functions satisfy the deformation equation \eqref{DefEqu} with $m=4$. Provided that the conditions of Theorem \ref{DefL2W} are fulfilled, the eigenvalues of \eqref{AddRS} solve the partial differential equation
\begin{equation*}
\sum_{k=1}^4 f_k(\lambda;u_1,u_2,u_3,u_4)\pD{\lambda}{u_k} = g(\lambda;u_1,u_2,u_3,u_4)
\end{equation*}

\subsection{Linear $2\times 2$ systems of the form $z'(x) = \big(\frac{1}{x}A+xB+C\big)z(x)$ on $(0,\infty)$}\subsep

Differential systems of this type have a regular singular point at $x=0$ and an irregular singularity at infinity. Again, if we assume that the condition \eqref{CTM} holds for $A,B\in\MzR$ and that $C\in\MzR$ satisfies $\tr C=0$, then we can transform such a system into
\begin{equation}
J y'(x) = \left(\lambda\begin{pmatrix} x & 0 \\[1ex] 0 & \frac{1}{x} \end{pmatrix} +
\frac{1}{x}\begin{pmatrix} 0 & \alpha \\[1ex] \alpha & \gamma \end{pmatrix} + 
x\begin{pmatrix} -\gamma & \beta \\[1ex] \ms\beta & 0 \end{pmatrix} + 
\begin{pmatrix} u_1 & u_2 \\[1ex] u_2 & u_3 \end{pmatrix}\right)y(x), \quad x\in(0,\infty)
\label{SysNR}
\end{equation}
by applying Theorem \ref{SysCTF} with $p(x)=\frac{1}{x}$ and $q(x)=x$. Apart from the values $\alpha,\beta>0$ and $\gamma\in\R$, which are supposed to be fixed, and the eigenvalue parameter $\lambda$ there are three more parameters $u_1,u_2,u_3$. We finally arrive at a linear Hamiltonian system $J y' = \left(\lambda W(x) + H(x)\right)y$ with coefficient matrices
\begin{gather*}
W(x) := \begin{pmatrix} x & 0 \\[1ex] 0 & \frac{1}{x} \end{pmatrix},\quad
H(x;u_1,u_2,u_3) := 
\frac{1}{x}\begin{pmatrix} 0 & \alpha \\[1ex] \alpha & \gamma \end{pmatrix} 
+ x \begin{pmatrix} -\gamma & \beta \\[1ex] \ms\beta & 0 \end{pmatrix} + 
\begin{pmatrix} u_1 & u_2 \\[1ex] u_2 & u_3 \end{pmatrix}
\end{gather*}
Now let us define the scalar functions
\begin{align*}
f_1 & := (2\alpha+1)(\lambda-\gamma)\lambda + u_1(\beta u_1-\alpha u_3+(\gamma-\lambda)u_2)\\
f_2 & := (\lambda^2-\gamma^2+\beta-u_1 u_3)\lambda \\
f_3 & := 2\beta(\lambda+\gamma)\lambda - u_3(\beta u_1-\alpha u_3+(\gamma+\lambda)u_2) \\
  g & := (\lambda^2-\gamma^2)u_2 + \alpha(\gamma-\lambda)u_3 - \beta(\gamma+\lambda)u_1
\end{align*}
and the differentiable matrix function
\begin{equation*}
G := \begin{pmatrix} 
\lambda(\lambda-\gamma) x & \beta\lambda x + \frac{1}{2}((\lambda-\gamma) u_2 - \beta u_1 + \alpha u_3) \\[1ex] \beta\lambda x + \frac{1}{2}((\lambda-\gamma) u_2 - \beta u_1 + \alpha u_3) & \lambda u_3 \end{pmatrix} 
\end{equation*}
These functions satisfy the deformation equation \eqref{DefEqu} with $m=4$, and if the remaining conditions of Theorem \ref{DefL2W} are also fulfilled, then the eigenvalues associated to \eqref{SysNR} solve the partial differential equation
\begin{equation*}
\sum_{k=1}^3 f_k(\lambda;u_1,u_2,u_3)\pD{\lambda}{u_k} = g(\lambda;u_1,u_2,u_3)
\end{equation*}
Here we won't examine all conditions required by Theorem \ref{DefL2W}, but we will at least determine the eigenvalues of the unperturbed differential operator generated by the differential expression $\tau y := W(x)^{-1}(J y' - H(x))y$ in the case $u_1=u_2=u_3=0$.

\begin{Lemma} \label{Laguerre}
For $u_1=u_2=u_3=0$ the differential system \eqref{SysNR} becomes
\begin{equation}
J y'(x) = \left(
\frac{1}{x}\begin{pmatrix} 0 & \alpha \\[1ex] \alpha & \lambda+\gamma \end{pmatrix} + 
x\begin{pmatrix} \lambda-\gamma & \beta \\[1ex] \beta & 0 \end{pmatrix}\right)y(x) \label{SysNR0}
\end{equation}
In case of $\alpha,\beta>0$ and $\gamma\in\R$ it has a nontrivial solution $y\not\equiv 0$ satisfying
\begin{equation}
\int_0^\infty y(x)\A\begin{pmatrix} x & 0 \\[1ex] 0 & \frac{1}{x} \end{pmatrix}y(x)\dx<\infty \label{IntNR0}
\end{equation}
if and only if $\lambda^2 = \gamma^2+4(\alpha+n+1)\beta$ with some non-negative integer $n$. Moreover, the associated eigenfunctions are given by
\begin{equation*}
y(x) = C\,\E^{-\frac{\beta}{2}x^2}x^\alpha\begin{pmatrix} 
2(\alpha+1+n)L_n^{(\alpha)}(\beta x^2) \\[1ex] (\gamma-\lambda)x^2 L_n^{(\alpha+1)}(\beta x^2) 
\end{pmatrix}
\end{equation*}
with some constant $C\in\C$, where $L_n^{(\alpha)}(z)$ denotes the generalized Laguerre polynomial of degree $n$.
\end{Lemma}

\begin{Proof} If we decompose $y(x)$ to its components $u(x)$ and $v(x)$, then we can write \eqref{SysNR0} as a system of coupled differential equations
\begin{align}
u'(x) & = (\tfrac{\alpha}{x}+\beta x)u(x)+\tfrac{\gamma+\lambda}{x}\,v(x) \label{NonRegU} \\
v'(x) & = (\gamma-\lambda) x u(x)-(\tfrac{\alpha}{x}+\beta x)v(x) \label{NonRegV}
\end{align}
Moreover, \eqref{IntNR0} holds if and only if both conditions
\begin{equation}
\int_0^\infty x\,u(x)^2\dx<\infty\quad\mbox{and}\quad\int_0^\infty \tfrac{1}{x}\,v(x)^2\dx<\infty \label{ConNR0}
\end{equation}
are fulfilled. In the case $\lambda=\gamma$ equation \eqref{NonRegV} implies $v(x)=C_2 x^{-\alpha}\exp(-x^2\beta/2)$ with some constant $C_2$. Since $\alpha>0$, the second integral in \eqref{ConNR0} is finite only for $C_2=0$, which means $v(x)\equiv 0$, and from equation \eqref{NonRegU} we get 
$u(x)=C_1 x^{\alpha}\exp(x^2\beta/2)$ with some constant $C_1$. To make the first integral in \eqref{ConNR0} finite, $C_1=0$ must apply because of $\beta>0$. Hence, if $\lambda=\gamma$, then the only solution which satisfies condition \eqref{IntNR0} is $y(x)\equiv 0$. We may therefore assume $\lambda-\gamma\neq 0$ and solve equation \eqref{NonRegV} for $u(x)$:
\begin{equation} \label{LaguerreU}
u(x) = \frac{x v'(x)+(\alpha+\beta x^2)v(x)}{x^2(\gamma-\lambda)}
\end{equation}
If we replace $u(x)$ in \eqref{NonRegU} with this term, then we receive for $v(x)$ the second order ODE
\begin{equation} \label{LaguerreV}
x^2 v''(x) - x\,v'(x) - \big(\beta^2 x^4+(2\alpha\beta+\gamma^2-\lambda^2)x^2+\alpha^2+2\alpha\big)v(x) = 0
\end{equation}
After applying the transformation $v(x) = w(\beta x^2)$, we finally obtain the Whittaker equation
\begin{equation*}
w''(z) + \left(-\frac{1}{4}+\frac{\kappa}{z} + \frac{\frac{1}{4}-\mu^2}{z^2}\right)w(z) = 0
\end{equation*}
for the function $w(z)$ with $z=\beta x^2\in(0,\infty)$, where 
\begin{equation*}
\kappa=\frac{\lambda^2-\gamma^2-2\alpha\beta}{4\beta},\quad \mu = \frac{\alpha+1}{2}
\end{equation*}
It has two fundamental solutions with the asymptotic behavior $w_1(z) = W_{\kappa,\mu}(z)\sim z^{\kappa}\E^{-z/2}$ and $w_2(z) = W_{-\kappa,\mu}(-z)\sim (-z)^{-\kappa}\E^{z/2}$ for $z\to\infty$ (see \cite[13.14.21 and 13.14.22]{NIST:2010}). The solutions of \eqref{LaguerreV}, for which $v(x)^2/x$ is integrable at $\infty$, must be constant multiples of $W_{\kappa,\mu}(\beta x^2)$. If $1/2-\kappa+\mu$ is \emph{not} a non-negative integer, then this fundamental solution behaves like $W_{\kappa,\mu}(\beta x^2)\sim C x^{-\alpha}$ for $x\to 0$ with some constant $C\neq 0$ (cf. \cite[13.14.15 and 13.14.16 with $1/2-\mu=-\alpha$]{NIST:2010}), so that $v(x)^2/x$ is not integrable at $0$. Therefore, in order to get a non-trivial solution of \eqref{SysNR0} which satisfies \eqref{IntNR0}, $\kappa = \mu+1/2+n$ must be valid with some non-negative integer $n$, and this means $\lambda^2 = \gamma^2+4(\alpha+1+n)\beta$. In this case the Whittaker functions can be written as generalized Laguerre polynomials (see \cite[13.18.17 with $\alpha+1$ instead of $\alpha$]{NIST:2010}):
\begin{equation*}
W_{\frac{\alpha+1}{2}+\frac{1}{2}+n,\frac{\alpha+1}{2}}(z) 
= (-1)^n n!\,\E^{-\frac{z}{2}} z^{\frac{\alpha+2}{2}}L_n^{(\alpha+1)}(z)
\end{equation*}
If we replace $z=\beta x^2$ and multiply by an appropriate factor, then
\begin{equation*}
v(x) = (\gamma-\lambda)\,\E^{-\frac{\beta}{2}x^2} x^{\alpha+2} L_n^{(\alpha+1)}(\beta x^2)
\end{equation*}
is a fundamental solution of \eqref{LaguerreV}, which satisfies $\int_0^\infty \frac{1}{x}v(x)^2\dx<\infty$. Finally, inserting $v(x)$ into \eqref{LaguerreU} and evaluating this expression gives
\begin{equation*}
u(x) = 2(\alpha+1+n)\,\E^{-\frac{\beta}{2}x^2} x^{\alpha}L_n^{(\alpha)}(\beta x^2)
\end{equation*}
by means of the differentiation formula \cite[18.9.24]{NIST:2010} and the recurrence relation \cite[18.9.14]{NIST:2010}.
\end{Proof}

\subsection{Linear $2\times 2$ systems $z'(x) = \big(\frac{1}{x}A+\Polynom\big)z(x)$ on $(0,\infty)$}\subsep

The system which has been studied in the last section is a special case of the more general system
\begin{equation*}
z'(x) = \left(\tfrac{1}{x}\,A + x^{m+1} B + \sum_{k=0}^m x^k C_k\right)z(x),\quad x\in (0,\infty) \end{equation*}
where we now replace the linear part $xB+C$ by a polynomial of degree $m+1$. Like \eqref{SysNR}, it has a regular singular point at $x=0$ and an irregular singularity at $\infty$. As in all our examples, we suppose that the matrices $A,B\in\MzR$ satisfy condition \eqref{CTM}, and we may also assume $\tr C_k=0$ $(k=0,\ldots,m)$ without restriction. By transformation into the complementary triangular form we obtain the linear Hamiltonian system
\begin{equation}\begin{split}
Jy'(x) = \left(\lambda W(x) + H(x)\right)y(x)\quad\mbox{with}\quad
W(x) := \begin{pmatrix} x^{m+1} & 0 \\[1ex] 0 & \frac{1}{x} \end{pmatrix}\quad\mbox{and} \\
H(x) := \tfrac{1}{x}\begin{pmatrix} 0 & \alpha \\[1ex] \alpha & \gamma \end{pmatrix} + 
x^{m+1}\begin{pmatrix} -\gamma & \beta \\[1ex] \ms\beta & 0 \end{pmatrix} + 
\sum_{k=0}^m x^k \begin{pmatrix} u_{3k+1} & u_{3k+2} \\[1ex] u_{3k+2} & u_{3k+3} \end{pmatrix} \label{PolNR}
\end{split}\end{equation}
Here $\alpha,\beta>0$ and $\gamma\in\R$ are supposed to be constant values, while $\lambda\in\R$ is considered to be the eigenvalue parameter. In addition, \eqref{PolNR} contains $3m+3$ (free) parameters $u_1,u_2,\ldots, u_{3m+3}$, and we can write this linear system in the form $\tau y = \lambda y$ with $\tau y := W^{-1}(J y'- H y)$. It should be mentioned that the eigenvalues in the unperturbed case $u_1 = \ldots = u_{3m+3} = 0$ can be obtained in a similar way as in Lemma \ref{Laguerre} by transforming the differential system to an appropriate Whittaker equation. Moreover, we define the scalar functions 
\begin{align*}
f_1 & := -Q u_1+(2\alpha+1)\lambda(\gamma-\lambda) \\
f_2 & := (\gamma^2-\lambda^2-\beta+u_1 u_{3m+3})\lambda \\
f_3 & := Q u_3-2\lambda\big(\beta(\gamma+\lambda)-u_2 u_{3m+3}\big) \\
  g & := \beta(\gamma+\lambda)u_{3m+1}+(\gamma^2-\lambda^2)u_{3m+2}+\alpha(\lambda-\gamma)u_{3m+3}
\end{align*}
where $Q := \beta u_{3m+1}-\alpha u_{3m+3}+(\gamma-\lambda) u_{3m+2}$, and
\begin{align*}
f_{3k+1} & := -Q u_{3k+1} + 2\lambda(\beta u_{3k-2}+(\gamma-\lambda)u_{3k-1}) \\
f_{3k+2} & := ((\gamma-\lambda)u_{3k}+u_{3k+1}u_{3m+3})\lambda \\
f_{3k+3} & := Q u_{3k+3} - 2\lambda(\beta u_{3k} - u_{3k+2}u_{3m+3})
\end{align*}
for $k=1,\ldots,m$. Now, if we introduce the matrix function
\begin{equation*}
G(x;\lambda;u_1,u_2,\ldots,u_{3m+3}) := \begin{pmatrix} 
\lambda(\gamma-\lambda)x & \frac{1}{2} Q - \beta\lambda x \\[1ex]
\frac{1}{2} Q - \beta\lambda x & -\lambda u_{3m+3}
\end{pmatrix}
\end{equation*}
then once more a lengthy calculation shows that the deformation equation
\begin{equation*}
\pD{G}{x} + (\lambda W + H)JG - G J(\lambda W + H) = \sum_{k=1}^{3m+3} f_k\pD{H}{u_k} + gW
\end{equation*}
is fulfilled, and the associated partial differential equation for the eigenvalues reads
$\sum_{k=1}^{3m+3} f_k\pD{\lambda}{u_k} = g$.

\subsection{An example with a linear $4\times 4$ Hamiltonian system}

In the previous examples we have only dealt with $2\times 2$ systems. However, the results from section \ref{sec:HamiltonSys} are valid for general $2n\times 2n$ systems of Hamiltonian type. We therefore want to conclude our studies with at least one such higher-dimensional example. We consider the linear $4\times 4$ system
\begin{gather*}
J y'(x) = (\lambda W(x) + H(x))y(x),\quad\mbox{where}\quad
W(x) := \begin{pmatrix} 
\frac{1}{1-x} & 0 & 0 & 0 \\[1ex]
0 & \frac{1}{1-x} & 0 & 0 \\[1ex]
0 & 0 &\ \frac{1}{x}\ & 0 \\[1ex]
0 & 0 & 0 &\ \frac{1}{x}\  
\end{pmatrix}\quad\mbox{and} \\
H(x) := \frac{1}{x}\begin{pmatrix}
0 & 0 & \alpha & 0 \\[1ex]
0 & 0 & 0 & \alpha \\[1ex]
\alpha & 0 & 0 & 0 \\[1ex]
0 & \alpha & 0 & 0 \end{pmatrix}
+ \frac{1}{1-x}\begin{pmatrix}
0 & 0 & \alpha & 0 \\[1ex]
0 & 0 & 0 & \alpha \\[1ex]
\alpha & 0 & 0 & 0 \\[1ex]
0 & \alpha & 0 & 0 \end{pmatrix} + \begin{pmatrix}
u_1+u_3 & u_1-u_3 & u_2+u_4 & u_2-u_4 \\[1ex]
u_1-u_3 & u_1+u_3 & u_2-u_4 & u_2+u_4 \\[1ex]
u_2+u_4 & u_2-u_4 & u_1+u_3 & u_1-u_3 \\[1ex]
u_2-u_4 & u_2+u_4 & u_1-u_3 & u_1+u_3 \end{pmatrix}
\end{gather*}
with some constant $\alpha>0$ and four parameters $u_1,\ldots,u_4\in\R$. In order to satisfy the deformation equation, we define the scalar functions
\begin{align*}
f_1 & := (2\lambda u_2 + 2 u_1 u_2 + u_1)(2\alpha u_3 - \lambda u_4) \\
f_2 & := (2\lambda u_1 + 2 u_1^2   + u_2)(2\alpha u_3 - \lambda u_4) \\
f_3 & := (2\lambda u_4 + 2 u_3 u_4 + u_3)(2\alpha u_1 - \lambda u_2) \\
f_4 & := (2\lambda u_3 + 2 u_3^2   + u_4)(2\alpha u_1 - \lambda u_2) \\
  g & := 2(2\alpha u_1 - \lambda u_2)(2\alpha u_3 - \lambda u_4)
\end{align*}
and the matrix function
\begin{equation*}
G(x;\lambda;u_1,u_2,u_3,u_4) := \begin{pmatrix}
x q_1 & x q_2 & (2x-1)q_3 & (2x-1)q_4 \\[1ex]
x q_2 & x q_1 & (2x-1)q_4 & (2x-1)q_3 \\[1ex]
(2x-1)q_3 & (2x-1)q_4 & (x-1)q_1 & (x-1)q_2 \\[1ex]
(2x-1)q_4 & (2x-1)q_3 & (x-1)q_2 & (x-1)q_1 \end{pmatrix}
\end{equation*}
which contains the factors
\begin{alignat*}{3}
q_1 & := 4\alpha u_1 u_3 - \lambda(u_2 u_3 + u_1 u_4), & \quad q_2 & := \lambda(u_2 u_3 - u_1 u_4) \\
q_3 & := \alpha(u_2 u_3 + u_1 u_4) - \lambda u_2 u_4,  & \quad q_4 & := \alpha (u_2 u_3 - u_1 u_4)
\end{alignat*}
These functions satisfy the deformation equation \eqref{DefEqu} with $m=4$, and assuming that the remaining requirements of Theorem \ref{DefL2W} are also fulfilled, the eigenvalues of this $4\times 4$ system solve the partial differential equation
\begin{equation*}
\sum_{k=1}^4 f_k(\lambda;u_1,u_2,u_3,u_4)\pD{\lambda}{u_k} = g(\lambda;u_1,u_2,u_3,u_4)
\end{equation*}

\section{Summary}

Our intention was to find a relationship between the eigenvalues of a linear $2n\times 2n$ Hamiltonian system $Jy' = (\lambda W+H)y$ and the entries of its coefficient matrix $H$. This problem has been solved to a certain extent. In section \ref{sec:HamiltonSys} we developed a method to derive a quasilinear first-order PDE for the eigenvalues $\lambda=\lambda(u_1,\ldots,u_m)$ provided that the coefficient matrix $H=H(x;u_1,\ldots,u_m)$ depends on several parameters and that it solves the deformation equation \eqref{DefEqu} or \eqref{DefLin}. Furthermore, in section \ref{sec:CompTriang} we investigated the problem how to convert an arbitrary $2\times 2$ differential system $z'(x)=\Phi(x)z(x)$ into a linear Hamiltonian system. With this theoretical background we were able to provide eigenvalue PDEs for a variety of boundary value problems in sections \ref{sec:RegSingular} and \ref{sec:NonRegular}. Even though most of these partial differential equations are rather complicated, they might be of practical use. In particular, such a PDE can be used to obtain a series expansion for the eigenvalues without having to calculate the associated eigenfunctions. As an example, we consider once more the differential operator generated by the linear Hamiltonian system \eqref{GCHE} with $\beta = \alpha > 0$ and $\gamma=0$, i.\,e., 
\begin{equation*}
\begin{pmatrix} 0 & -1 \\[1ex] 1 & \ms 0 \end{pmatrix}y'(x) 
= \left(\lambda\begin{pmatrix} \frac{1}{1-x} & 0 \\[1ex] 0 & \frac{1}{x} \end{pmatrix}
+ \begin{pmatrix} u_1 & \frac{\alpha}{x}+\frac{\alpha}{1-x}+u_2 \\[1ex] \frac{\alpha}{x}+\frac{\alpha}{1-x} + u_2 & u_3 \end{pmatrix}\right)y(x),\quad x\in(0,1)
\end{equation*}
According to Theorem \ref{EigRS} the eigenvalues $\lambda=\lambda(u_1,u_2,u_3)$ depend analytically on $(u_1,u_2,u_3)\in\Dom$ in some domain $\Dom\subset\R^3$ with $(0,0,0)\in\Dom$, where 
\begin{equation} \label{Ansatz}
\lambda(u_1,u_2,u_3) = \lambda_0 + c_1 u_1 + c_2 u_2 + c_3 u_3 + \sum_{i+j+k\geq 2}^\infty c_{i,j,k} u_1^i u_2^j u_3^k
\end{equation}
and $\lambda_0 := \lambda(0,0,0) = \pm(2\alpha+n+1)$ with some non-negative integer $n$. In order to get a linear approximation for the eigenvalues with respect to the parameters, we have to calculate the coefficients $c_1,c_2,c_3$. In case of $\beta=\alpha$ and $\gamma=0$ the partial differential equation \eqref{PDERS} reads
\begin{equation}\begin{split}
& (\lambda u_1 + 2\lambda^2 u_2 - \alpha u_1^2 + \lambda u_1 u_2 + \alpha u_1 u_3)\pD{\lambda}{u_1} 
+ (\lambda^2 u_1 + \lambda u_2 + \lambda^2 u_3 + \lambda u_1 u_3)\pD{\lambda}{u_2} + {} \\ & \quad {} 
+ (2\lambda^2 u_2 + \lambda u_3 + \alpha u_1 u_3 + \lambda u_2 u_3 - \alpha u_3^2)\pD{\lambda}{u_3} 
- (\alpha\lambda u_1 - \lambda^2 u_2 + \alpha\lambda u_3) = 0 \label{PDE3}
\end{split}\end{equation}
By inserting \eqref{Ansatz} into \eqref{PDE3} we obtain
\begin{equation*}
0 = \lambda_0(c_1 + \lambda_0 c_2 - \alpha)u_1
  + \lambda_0(2\lambda_0 c_1 + c_2 + 2\lambda_0 c_3 + \lambda_0)u_2 
  + \lambda_0(\lambda_0 c_2 + c_3 - \alpha)u_3 + \mbox{higher order terms}
\end{equation*}
and setting the prefactors to zero gives
\begin{equation*}
c_1 = c_3 = -\frac{\lambda_0^2 + \alpha}{4\lambda_0^2 - 1},\quad c_2 = \frac{(4\alpha+1)\lambda_0}{4 \lambda_0^2 - 1}
\end{equation*}
If we ignore the terms of total degree ${}>1$ in $u_1,u_2,u_3$, then we get the linear approximation
\begin{equation*}
\lambda(u_1,u_2,u_3) \approx \lambda_0 - \frac{\lambda_0^2+\alpha}{4\lambda_0^2-1}\,u_1 + 
\frac{(4\alpha+1)\lambda_0}{4\lambda_0^2-1}\,u_2 - \frac{\lambda_0^2+\alpha}{4\lambda_0^2-1}\,u_3
\quad\mbox{as}\quad (u_1,u_2,u_3)\to(0,0,0)
\end{equation*}
Now, let us consider the case $\alpha=2$, where $n=0$ produces the lowest positive eigenvalue of the unperturbed operator $\lambda(0,0,0)=+(2\alpha+n+1)=5$. For a small perturbation, say $u_1=0.25$, $u_2=0.2$ and $u_3=-0.05$, the linear approximation predicts $\lambda(0.25,0.2,-0.05)\approx 5.0363636$ in good agreement with the numerically calculated eigenvalue $5.0367231$, which has been computed up to eight decimal digits by means of an appropriate shooting method. If we include also quadratic terms in the approximation of \eqref{Ansatz} without doing the calculation here in detail, then we get a better estimate $\lambda(0.25,0.2,-0.05)\approx 5.0367372$. By this method, which is only briefly sketched here, one can determine a series expansion for the eigenvalues of other linear Hamiltonian systems as well. However, this is a task that exceeds the scope of the present paper.

\end{document}